\documentclass[traditabstract,longauth]{aa}
\usepackage{graphicx}
\usepackage{txfonts}
\usepackage{hyperref}
\usepackage[svgnames]{xcolor}
\usepackage{amsmath}
\usepackage{natbib}
\usepackage{lscape}  
\usepackage{multirow}

\hypersetup{
     colorlinks   = True,
     citecolor    = Blue,
     linkcolor = Blue,
     urlcolor = DarkBlue
}

\newcommand{\add}[1]{{\color{black}#1}}

\newcommand{\lya}{Ly$\alpha$ }
\newcommand{\ciii}{C\textsc{III}]$\lambda$1909 }

\begin{document}

   \title{The VIMOS Ultra-Deep Survey: the Ly$\alpha$ emission line morphology at $2<z<6$\thanks{Based on data obtained with the European
          Southern Observatory Very Large Telescope, Paranal, Chile, under Large
          Program 185.A--0791. }}

\author{
B. Ribeiro \inst{1,2}
\and O.~Le F\`evre\inst{2}
\and A.~Paulino-Afonso\inst{3}
\and P.~Cassata\inst{4,5}
\and V.~Le Brun\inst{2}
\and B.~C.~Lemaux \inst{6}
\and D.~Maccagni\inst{7}
\and L.~Pentericci\inst{8}
\and R.~Thomas\inst{9}
\and G.~Zamorani \inst{11}
\and E.~Zucca\inst{11}
\and R.~Amor\'in\inst{12,13}
\and S.~Bardelli\inst{11}
\and L.~P.~Cassar\`a\inst{7}
\and L.~ Guaita\inst{14}
\and N.P.~Hathi\inst{15}
\and A.~Koekemoer\inst{15}
\and D. Schaerer\inst{16,17}
\and M.~Talia\inst{10,11}
\and J.~Pforr\inst{18}
\and L.~Tresse\inst{19,2}
\and S.~Fotopoulou\inst{20}
\and D.~ Vergani\inst{11}}

\institute{Leiden Observatory, Leiden University, NL-2300 RA Leiden, Netherlands
\and
Aix Marseille Univ., CNRS, CNES, LAM, Marseille, France
\and
CENTRA - Centro de Astrof\'isica e Gravita\c{c}\~ao, Instituto Superior T\'ecnico, Av. Rovisco Pais, 1, P-1049-001 Lisboa, Portugal
\and
Dipartimento di Fisica e Astronomia, Universit\`a di Padova, Vicolo dell'Osservatorio, 3 35122 Padova, Italy
\and
INAF Osservatorio Astronomico di Padova, vicolo dell'Osservatorio 5, I-35122 Padova, Italy
\and 
Department of Physics, University of California, Davis, One Shields, Ave., Davis, CA 95616, USA
\and
INAF--IASF Milano, via Bassini 15, I--20133, Milano, Italy
\and
INAF--Osservatorio Astronomico di Roma, via di Frascati 33, I-00040, Monte Porzio Catone, Italy
\and
European Southern Observatory, Av. Alonso de Córdova 3107, Vitacura, Santiago, Chile
\and
University of Bologna - Department of Physics and Astronomy, Via Gobetti 93/2, I-40129, Bologna, Italy 
\and
 INAF - Osservatorio di Astrofisica e Scienza dello Spazio, Via Gobetti 93/3, I-40129, Bologna, Italy\and
Instituto de Investigaci\'on Multidisciplinar en Ciencia y Tecnolog\'ia, Universidad de La Serena, Ra\'ul Bitr\'an 1305, La Serena, Chile
\and
Departamento de F\'isica y Astronom\'ia, Universidad de La Serena, Av. Juan Cisternas 1200 Norte, La Serena, Chile
\and
N\'ucleo de Astronom\'ia, Facultad de Ingenier\'ia, Universidad Diego Portales, Av. Ej\'ercito 441, Santiago, Chile
\and
Space Telescope Science Institute, 3700 San Martin Drive, Baltimore, MD 21218, USA
\and
Geneva Observatory, University of Geneva, ch. des Maillettes 51, CH-1290 Versoix, Switzerland
\and
Institut de Recherche en Astrophysique et Plan\'etologie - IRAP, CNRS, Universit\'e de Toulouse, UPS-OMP, 14, avenue E. Belin, F31400, Toulouse, France
\and
ESA/ESTEC SCI-S, Keplerlaan 1, 2201 AZ, Noordwijk, The Netherlands
\and
Univ Lyon, Univ Lyon 1, CNRS, Centre de Recherche Astrophysique de Lyon UMR5574, F-69230, Saint-Genis-Laval, France
\and
Centre for Extragalactic Astronomy, Department of Physics, Durham University, South Road, Durham DH1 3LE, U.K.
 \\ \\
             \email{ribeiro@strw.leidenuniv.nl}
}

   \date{}


  \abstract
   {The Lyman-$\alpha$ (Ly$\alpha$) emission line has been ubiquitously used to confirm and study high redshift galaxies. We report on the line morphology as seen in the 2D spectra from the VIMOS Ultra Deep Survey (VUDS) in a sample of 914 \lya\ emitters {from a parent sample of 4192 star-forming galaxies at} $2<z_\mathrm{spec}\lesssim6$. The study of the spatial extent of \lya\ emission provides insight into the escape of Ly$\alpha$ photons from galaxies. We classify the line emission as either non-existent, coincident, \add{projected spatial} offset, or extended with respect to the observed 2D UV continuum emission. The line emitters in our sample are classified as $\sim45\%$ coincident, $\sim24\%$ extended and $\sim11\%$ offset emitters.  \add{For galaxies with detected UV continuum,} we show that extended Ly$\alpha$ emitters (LAEs) correspond to the highest equivalent width galaxies (with an average $W_\mathrm{Ly\alpha}\sim-22$\AA). This means that this class of objects is the most common in narrow-band selected samples, which usually select high equivalent width LAEs, $<-20$\AA. Extended Ly$\alpha$ emitters are found to be less massive, less star-forming, with lower dust content, and smaller UV continuum sizes ($r_{50}\sim0.9$kpc) \add{of all the classes considered here}. We also find that galaxies with larger UV-sizes have lower fractions of Ly$\alpha$ emitters. By stacking the spectra per emitter class we find that the weaker Ly$\alpha$ emitters have stronger low ionization inter-stellar medium (ISM) absorption lines. Interestingly, we find that galaxies with Ly$\alpha$ offset emission ({median separation of} $1.1_{-0.8}^{+1.3}$kpc from UV continuum) show similar velocity offsets in the ISM as those with no visible emission (and different from other Ly$\alpha$ emitting classes). This class of objects may hint at episodes of gas accretion, bright offset clumps or on-going merging activity into the larger galaxies.
 \vspace{+0.0cm}
}

   \maketitle
%

\section{Introduction}\label{sec:intro}

The Lyman-$\alpha$ (Ly$\alpha$) emission line is the intrinsically brightest feature in a hot source spectrum \citep[e.g.][]{schaerer2003}. Combined with the fact that {this line is redshifted} into the optical and near-infrared regime for a broad redshift range ($2<z\lesssim10$), it is one of the most used tracers of galaxies at high redshift. However, due to its resonant nature it is easily scattered by gas clouds (HI) and its emission can be suppressed by the presence of dust.

The radiative transfer process that eventually leads the Ly$\alpha$ photons to escape from their emitting region gives rise to a number of line profiles that have been already reported in observations: asymmetric line profiles \citep[e.g.][]{kunth1998,rhoads2003,shapley2003,shimasaku2006,tapken2007,vanzella2010,lidman2012,wofford2013}, double peak profiles \citep[e.g.][]{shapley2006,quider2009,kulas2012,yamada2012,matthee2018} as well as spatially extended emission surrounding galaxies \citep[e.g.][]{fynbo2001,steidel2003,steidel2011,rhoads2005,hayes2005,hayes2007,rauch2008,ostlin2009,matsuda2012,wisotzki2016}. In an attempt to explain such diversity of profiles theoretical and numerical models have been proposed \citep[e.g.][]{harrington1973,neufeld1990,neufeld1991,loeb1999,ahn2000,zheng2002,dijkstra2006,hansen2006,tasitsiomi2006,verhamme2006,verhamme2012,laursen2009,duval2014,gronke2015}.

{The detection of the \lya\ line is} a very successful method to identify high redshift galaxies using large narrow-band surveys in the optical \citep[e.g.][]{cowie1998, rhoads2000, shimasaku2006, ouchi2008, ouchi2010, ouchi2018, sobral2017, sobral2018, hao2018} and to spectroscopically detect or confirm high redshift candidates \citep[e.g.][]{martin2004, lefevre2005, cassata2011, ono2012, finkelstein2013, bacon2015, lefevre2015, zitrin2015} and large samples of such galaxies have been collected to date. From these samples, the study of their rest-frame UV morphologies at $z>2$ is also widely covered \citep[e.g.][]{pirzkal2007,taniguchi2009,bond2009,bond2011,bond2012, gronwall2011,kobayashi2016,paulino-afonso2017}. When considering the local Universe and due to the impossibility of obtaining rest-frame UV observations from ground observatories, the Ly$\alpha$ Reference Sample \citep[LARS,][]{ostlin2014} is the only study that focuses on the morphology of Ly$\alpha$ emitters  \citep[LAEs,][]{guaita2015}.

Ly$\alpha$ emission is often found to be more extended (in a diffuse halo) than the stellar UV continuum emission \citep[e.g.][]{rauch2008,finkelstein2011,steidel2011,matsuda2012,feldmeier2013,momose2014,momose2016,matthee2016,wisotzki2016,leclercq2017, xue2017}. The process responsible for such observations is thought to be the scattering of photons by neutral HI gas around galaxies at high redshift \citep[e.g.][]{zheng2011}. Galaxies with observed Ly$\alpha$ emission \add{are found to have small rest-frame UV sizes} at all observed redshifts \citep[e.g][]{venemans2005,malhotra2012,paulino-afonso2017}. Such findings are at odds with the stronger evolution in galaxy sizes seen for \add{galaxy populations selected on other criteria \citep[using similar size estimates on Lyman-break or UV continuum selected galaxies, especially at $z<4$, e.g.][]{ferguson2004,bouwens2004,vanderwel2014, morishita2014, shibuya2016,ribeiro2016}.} Additionally, {there is evidence} for a positive correlation between line luminosity and galaxy UV continuum size \citep[e.g.][]{hagen2014,paulino-afonso2017} as well as between the extent of Ly$\alpha$ with the extent of the UV emission \citep[e.g.][]{wisotzki2016, leclercq2017, xue2017}. { There also seems to be a correlation between galaxy UV size and the escape of ionizing radiation, with the smallest galaxies having the highest escape fraction \citep[e.g.][]{marchi2018}.}

The mechanisms through which Ly$\alpha$ photons escape into inter-galactic space are complex and photons can travel for several kpc in random walks before free-roaming towards our line of sight \citep[e.g.][]{zheng2011,dijkstra2012,rosdahl2012,lake2015,trebitsch2017,kimm2019}. Such large excursions increase the chance of dust absorption corresponding to photon destruction \citep[e.g.][]{neufeld1991,laursen2013}. And there are simulations that show that the escape fraction is linked to the galaxy inclination with respect to the line of sight \citep{verhamme2012,berhens2014}. Observationally, the escape fraction of Ly$\alpha$ photons is anti-correlated to dust attenuation and star formation rate (SFR) but large scatter is observed \citep[e.g.][]{hayes2010,hayes2011,atek2014,matthee2016}. Other studies show that a key quantity in determining Ly$\alpha$ emissivity is the column density of neutral hydrogen \citep[e.g.][]{shibuya2014a,shibuya2014b}.

The comparison of LAEs {(usually with $W_\mathrm{Ly\alpha} < -20\AA$)} with galaxies with weak or no Ly$\alpha$ emission is ubiquitous in the literature \citep[e.g][]{shapley2001,erb2006,gawiser2006,gawiser2007,pentericci2007,lai2008,reddy2008, finkelstein2009,kornei2010,nilsson2011,acquaviva2012,lefevre2015,cassata2015,hathi2016,oyarzun2017,santos2020}.  While some studies indicate that LAEs are {low mass, dust poor galaxies} and likely the early stage of galactic evolution \citep[e.g.][]{erb2006, gawiser2006,gawiser2007,pentericci2007,santos2020} others point to a scenario where LAEs can be older \citep[][]{kornei2010} and more massive \citep[e.g.][]{lai2008}. Some find small differences between LAEs and the typical star-forming population when drawn from the same parent selection\citep[e.g.][]{hathi2016}. \add{Relatively extreme optical emission line galaxies detected at $z\sim2$ ([OII] and [OIII] emitters with average EW([OIII])$\sim$200\AA) are found to have similar stellar masses to LAEs, but they are less massive and less star forming than the most massive color-selected star-forming galaxies at these redshifts \citep[e.g.][]{hagen2016}.}

In this paper we report a taxonomy study on the shape of the 2D Ly$\alpha$ emission line using data from the VIMOS Ultra Deep Survey \citep[VUDS,][]{lefevre2015}, a large spectroscopic survey with secure spectroscopic redshifts for $4192$ galaxies at $2\lesssim z \lesssim 6$ representative of the star-forming population at this epoch. We divide our sample into different classes according to the Ly$\alpha$ 2D shape with respect to the UV continuum and then report on the abundance of each class as a function of redshift and estimate median key physical parameters (stellar mass, SFR, age, dust extinction) associated with each class. {We also attempt at interpreting each class with an schematic 3D model of line emission. We use the acronym LAGs for Ly$\alpha$ emitting Galaxies (defined as objects with secure classification of visible Ly$\alpha$ emission from their 2D spectra) instead of LAEs to avoid confusion with the standard definition of LAEs in the literature.}

This paper is organized as follows. In Sect. \ref{sec:data_sample} we briefly describe the VUDS survey and we highlight our sample selection. In Sect. \ref{sec:lya_classification} we describe the classification scheme that we designed as well as the method used for the classification. In Sect. \ref{sec:lya_shapes} we summarize our findings in terms of the impact of redshift, stellar mass, line luminosity, line equivalent width, and UV luminosity on the derived classification. In Sect. \ref{sec:lya_physics} we present and discuss the physical properties associated with each class. We discuss our findings in the context of current radiative transfer models of Ly$\alpha$ in Sect. \ref{sec:discussion} and summarize the main results in Sect. \ref{sec:summary}. We use a cosmology with $H_0=70~\mathrm{km~s^{-1}~Mpc^{-1}}$,
$\Omega_{0,\Lambda}=0.7$ and $\Omega_{0,m}=0.3$. All magnitudes are given in the AB system \citep[][]{oke1983}. {Equivalent widths ($W$) are in the rest-frame with the convention that negative denotes emission. Line fluxes are in $\mathrm{erg\ s^{-1}cm^{-2}}$ units unless explicitly stated otherwise.}


\section{Data}\label{sec:data_sample}

As presented by \citet[][]{lefevre2015}, VUDS is a large spectroscopic survey that targeted $\sim10 000$ objects covering an area of {1 deg$^{2}$} on three separate fields: COSMOS, ECDFS, and VVDS-02h. With the objective to observe galaxies in the redshift range $2<z<6+$, targets were selected based on the first or second photometric redshift peaks being at $z_\mathrm{phot}+1\sigma>2.4$ \add{and typically at $i_{AB}\leq25$. In addition, a small number of targets were selected using a redshift specific} color-color criterion for being at a specific redshift \add{, e.g., a Lyman Break Galaxy (LBG) selection for $z\ge2$ galaxies, if not already selected through their photometric redshift.}. {Since for these we can use the set of $ugrizYJHK$ filters to estimate the break in the continuum, \add{the latter category of targets} are not necessarily brighter than $i_{AB} = 25$.} To fill in the remaining available space on the masks, randomly selected samples with $23 < i_{AB} < 25$ were added to the target list. The spectra were obtained using the VIMOS spectrograph on the ESO-VLT covering, with two low resolution grisms (R=230), a wavelength range of $3650\AA < \lambda < 9350\AA$. The total integration is $\sim 14$h per pointing and grism.

Data processing is performed within the VIPGI environment \citep{scoddeggio2005} and is followed by extensive redshift measurements campaigns using the EZ redshift measurement engine  \citep{garilli2010}. At the end of this process each galaxy has  flux and wavelength calibrated 2D and 1D spectra, a spectroscopic redshift measurement, and associated redshift reliability flag. For further information on the data processing we refer to \citet{lefevre2015}.


\begin{figure*}
\centering
\includegraphics[width=0.35\linewidth]{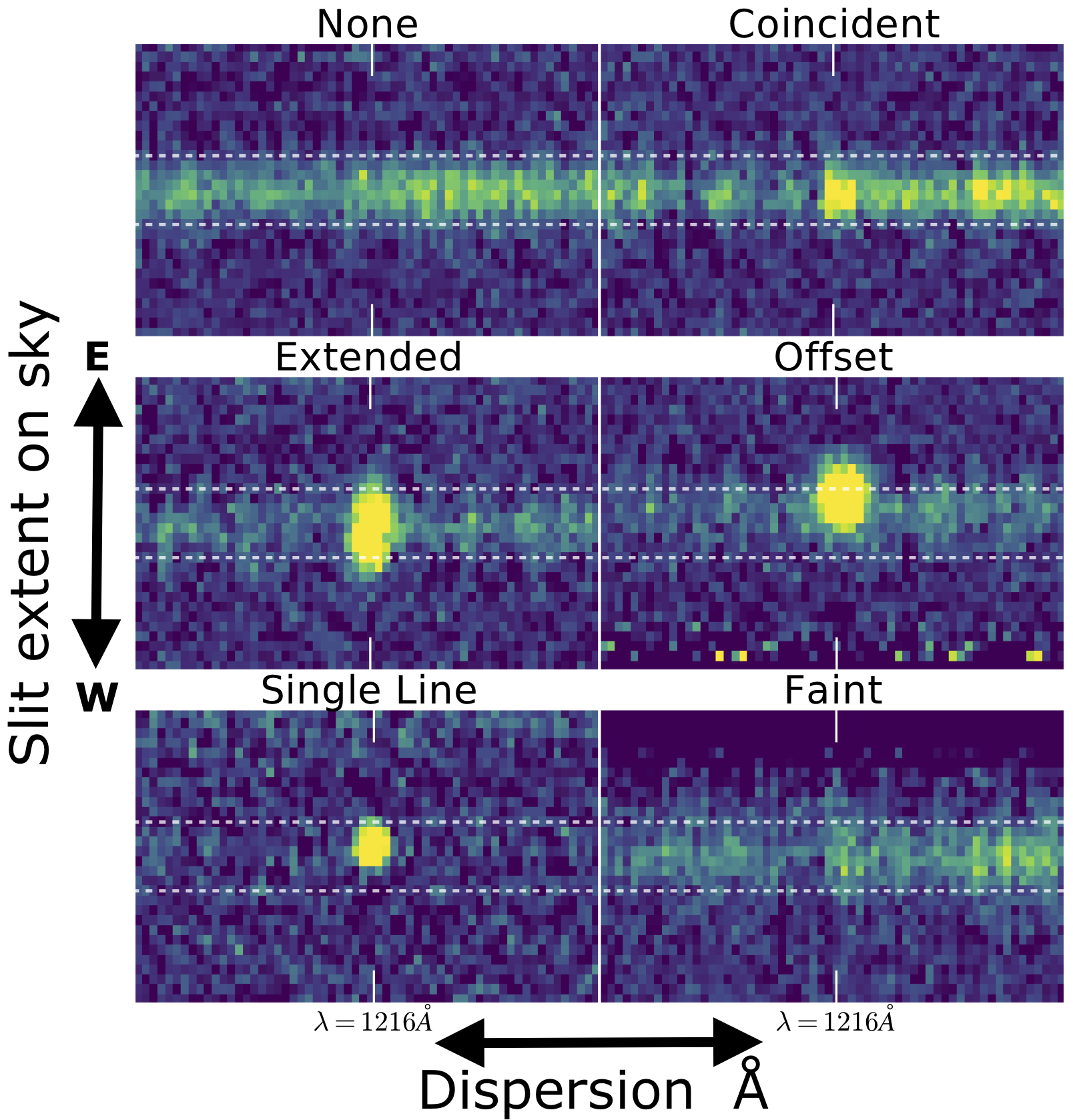}
\includegraphics[width=0.60\linewidth]{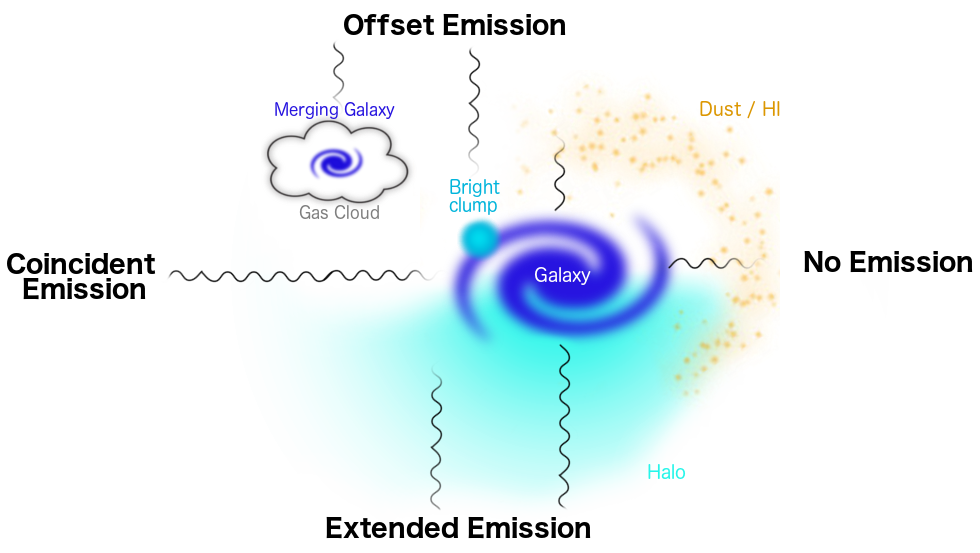}
\caption{ {\bf Left}: Examples of each of the main classes that are defined for the purpose of this study: galaxies with no discernible emission or with Ly$\alpha$ absorption (None); galaxies with emission of similar extent to the continuum (Coincident); lines with larger extent than the continuum (Extended); lines with offset emission with respect to the continuum (Offset); emission with no associated continuum (Single Line); possible, faint emission (Faint). The white dashed lines delimit a 1.64\arcsec\ area centred on the extraction window. {\bf Right}: schematic view of possible physical explanations for each observed class in the 2D spectra. }
\label{fig:class_examples}
\end{figure*}

To conduct this study we selected all primary and secondary target galaxies in the VUDS survey for which we have the most reliable redshift measurements, i.e. galaxies with redshift flags 2 and 9 ($\sim$80\% probability to be correct) and flags 3 and 4 (95 -100\% probability to be correct). For more details on the reliability flags we refer to \citet{lefevre2015}. This constraint led to a total of 4360 targets {which after removing duplicate objects results in a total of 4192 individual galaxies (2100 in COSMOS, 1575 in VVDS-02h and 517 in ECDFS)} to be visually inspected for the presence of the Ly$\alpha$ line and subsequently classified.

\subsection{Spectral Energy Distribution}

We use the physical parameters derived from the simultaneous SED fitting of the VUDS spectra and all multi-wavelength photometry available for each galaxy, using the code GOSSIP+ as described by \cite{thomas2016}. This method expands the classical SED fitting technique to make use of the UV rest-frame spectra in addition to photometry, improving the accuracy of physical parameter measurements \citep[for more details about the improvement, see][]{thomas2016}.  It also provides measurements of physical quantities such as galaxy ages as well as the IGM transmission along the line of sight of each galaxy, an improvement compared to using a fixed transmission at a given redshift \citep{thomas2015}.

To fit the photometric+spectroscopic data we use templates derived from \citet{bruzual2003} models with a \citet{chabrier2003} initial mass function. To model the star formation histories (SFH) we use  an exponentially declining parametrization as $\mathrm{SFR} \propto \exp{-t/\tau}$ ($\tau$ in the range 0.1 Gyr to 30 Gyr) and two additional models with delayed SFH peaking at 1 and
3 Gyr. The templates are created in a grid of 51 ages (in the range 0.1 Gyr to 14.5 Gyr). We apply
a \citet{calzetti2000} dust extinction law to the templates, using $E_s(B-V)$ in the range 0
to 0.5. Models with four different metallicities are used {($Z=0.004,0.008,0.02,0.05$)}.
The parameters of interest in this paper are the stellar mass (M$_{\star}$), $SFR$, dust extinction ($A_V$) and {age \citep[$t \equiv \mathcal{A}_\mathrm{onset}$ as described by][]{thomas2016}}, for which the median values of the probability density function are used. {We expect a statistical uncertainty of $\sim$0.1 dex in stellar masses, $\sim$0.15 dex in SFR, and a $\sim10\%$ uncertainty on the age \citep{thomas2016}. For further details about typical uncertainties on these quantities see also \citet[][]{ilbert2013} and \citet{tasca2015}.}

\subsection{Galaxy sizes}\label{ssection:galSizes}

We use GALFIT \citep[][]{peng2002,peng2010} on F814W images from COSMOS \citep{koekemoer2007} and CANDELS \citep{koekemoer2011} to derive effective radii and use the total extent measurement $r_T^{100}$ described by \citet{ribeiro2016} which defines the radius of a circle with the same area as the total extent of the galaxy above a limiting isophote. We also use our $r_T^{50}$ which is derived from the area containing 50\% of the galaxy light above the same isophote. For all size related analysis, we restrict our {sample of galaxies to have HST imaging coverage in the COSMOS and ECDFS fields}, to have stellar masses $\log(M_\star/M_\odot)>9.5$ and to be in the redshift range $2<z<4.5$ where our measurements of total extent are valid \citep[see][for more details]{ribeiro2016}. {This yields a total of 1040 galaxies with size measurements.}

\subsection{Ly$\alpha$ line properties}\label{ssec:lya_measures}
The line fluxes and equivalent widths were measured as described by \citet[][]{cassata2015}. In summary, we used the \textsc{noao.onedspec.splot} tool available from \textsc{iraf} \citep[][]{tody1986,tody1993}. The continuum was estimate from two manually chosen points, one on each side of the line. The flux of the line is then computed from direct integration of the spectra above the estimated continuum. This allows to estimate accurately the properties of individual lines which often deviate from a simple Gaussian in the case of Ly$\alpha$. Doing this process interactively also minimizes possible issues in the continuum estimation that may from defects in individual spectra. The uncertainties were computed using the formalism from \citet{tresse1999} and we estimate \add{a typical} uncertainty of $\pm5$\AA\ (for weak absorbers and emitters) and up to $\pm25$\AA\ (for strong absorbers and emitters).

We note, however, that these measurements were all performed on the extracted one-dimensional spectra. From experience in the survey we know that the spectral extraction window is often defined by the extent of the UV continuum on the 2D spectra and misses some of the flux of the Ly$\alpha$ line, especially in cases where the line is extended or even offset from the continuum (see e.g. Fig. \ref{fig:class_examples}, left panel). We compute the ratio of the line flux of initial extracted 1D spectra to the flux extracted in an aperture that is centred on and encompasses the full extent of the \lya\ line to estimate the level of flux loss in Ly$\alpha$. We find an average flux loss of $\sim$19\% for offset \lya\ emitters \add{(with 68\%  of the galaxies - calculated from the 16th and 84th percentiles - having flux losses between 5\% and 38\%)} and $\sim$8\% for extended \lya\ emitters \add{(with 68\% of the galaxies having flux losses between 3\% and 15\%)}. We do not apply any correction to the individual spectra since this level of flux loss estimate does not affect any conclusions from our study.

\section{Classification Scheme}\label{sec:lya_classification}

We use the 2D spectrograms produced by VIMOS slits, with the spatial dimension of the spectrogram covering on average 10\arcsec\ across the galaxy, and the spectral direction including the Ly$\alpha$1216\AA\ line location, as well as the continuum emission, up to or beyond the C{\sc iii}]$\lambda\lambda$1907,1909\AA\ emission \citep{lefevre2017}. Our classification scheme identifies the position of the {\lya\ line} in this 2D spectrogram. Such scheme aims to divide galaxies with \lya\ in emission with respect to the their spatial properties as seen in the 2D spectra that were obtained with the VIMOS spectrograph. Our main classes are:

\begin{itemize}
\item {\bf None}, where there is no line emission detected in the 2D spectra, or the line is observed in absorption.
\item {\bf Coincident}, where the total extent of the emission line coincides with the extent of the galaxy continuum.
\item {\bf Extended}, where the total extent of the emission line is more extended than that of the  galaxy UV continuum{, but still centred on the continuum in the spatial direction.}
\item {\bf Offset}, where there is {an offset in the spatial direction} between the center of the emission line and the center of the UV continuum emission.
\end{itemize}
There are two additional classes that we use in this classification scheme. {\bf Single Line} emitters, where there is only an emission line and no UV continuum flux is detected. {\bf Faint} emitters, where there is no clear detection of an emission line in the 2D spectra making it impossible for it to be properly classified as one of the four main classes. Examples of each of the classes are {shown in the left panel of Fig. \ref{fig:class_examples}.}

Each class can be schematically represented by different modes of Ly$\alpha$ escape from the galaxy. Coincident emission occurs when there is little to no scattering of the radiation and we observe the emission from its original regions. Extended emission is when there is significant scatter on the surrounding gas and we observe the halo around the galaxy. Offset emission can happen for one of two reasons, either there are two clumps/galaxies {spatially offset from each other and one is emitting \lya\ and the second is not} or there is a single galaxy with an asymmetric gas/dust distribution (potentially caused from stellar winds and/or supernovae explosions) from where the Ly$\alpha$ photons preferentially escape from one side of the galaxy. If no emission is observed there is likely high dust contents or high density of neutral hydrogen preventing Ly$\alpha$ from being observed in emission. Faint and single line emitters have low S/N in the line and continuum, respectively, and thus are purely observational classes. These are summarized {in the right panel of Fig. \ref{fig:class_examples}.}

\subsection{Classification procedure}\label{ssection:class_proc}

\begin{table*}
\centering
\caption{{Number of LAGs in each redshift bin and for each selection that we use in this paper.}}
\begin{tabular}{ccccccc}
\hline
 Selection & $2.0<z<2.5$ & $2.5<z<3.0$ & $3.0<z<3.5$ &  $3.5<z<4.5$ &  $4.5<z<6.5$ & Total \\
 \hline
All& 171 & 336 & 194 & 136 & 76 & 913* \\
\hline
$9.0<\log_{10}\left(M_\star/M_\odot\right)\leq9.5$& 44 & 103 & 49 & 29 & 9 & 234 \\
$9.5<\log_{10}\left(M_\star/M_\odot\right)\leq9.8$& 43 & 85 & 34 & 38 & 11 & 211 \\
$9.8<\log_{10}\left(M_\star/M_\odot\right)\leq10.1$& 31 & 52 & 48 & 27 & 9 & 167 \\
$10.1<\log_{10}\left(M_\star/M_\odot\right)\leq12.0$& 32 & 64 & 41 & 31 & 26 & 194 \\
\hline
$-10.0<W_\mathrm{Ly\alpha} [\mathrm{\AA}]\leq-5.0$& 31 & 66 & 29 & 12 & 6 & 144 \\
$-20.3<W_\mathrm{Ly\alpha} [\mathrm{\AA}]\leq-10.0$& 30 & 71 & 41 & 32 & 14 & 188 \\
$-46.2<W_\mathrm{Ly\alpha} [\mathrm{\AA}]\leq-20.3$& 46 & 79 & 38 & 39 & 15 & 217 \\
$-769.7<W_\mathrm{Ly\alpha} [\mathrm{\AA}]\leq-46.2$& 31 & 59 & 51 & 37 & 27 & 205 \\
\hline
$-19.8<\log_{10}\left(F_\mathrm{Ly\alpha}\right)\leq-17.2$& 8 & 25 & 16 & 10 & 10 & 69 \\
$-17.2<\log_{10}\left(F_\mathrm{Ly\alpha}\right)\leq-16.8$& 28 & 78 & 49 & 18 & 13 & 186 \\
$-16.8<\log_{10}\left(F_\mathrm{Ly\alpha}\right)\leq-16.4$& 46 & 90 & 48 & 36 & 17 & 237 \\
$-16.4<\log_{10}\left(F_\mathrm{Ly\alpha}\right)\leq-13.4$& 72 & 115 & 59 & 59 & 22 & 327 \\
\hline
$-20.3<M_{FUV}\leq-19.5$& 70 & 69 & 18 & 4 & 1 & 162 \\
$-20.7<M_{FUV}\leq-20.3$& 29 & 84 & 48 & 8 & 4 & 173 \\
$-21.1<M_{FUV}\leq-20.7$& 20 & 60 & 52 & 33 & 4 & 169 \\
$-23.5<M_{FUV}\leq-21.1$& 13 & 47 & 43 & 70 & 47 & 220 \\
\hline
\multicolumn{7}{l}{*{\footnotesize The missing galaxy that is not included in our analysis is a single line emitter at $z=6.5363$}.} \\
\end{tabular}
\label{tab:numbers}
\end{table*}

{In order to carry out the individual visual classification of the 4360 spectra (corresponding to 4192 individual galaxies)} a total of 10 collaborators inspected the data set. The sample was divided in 20 data packages of 218 targets each and then each volunteer was assigned 4 different data packages. We distributed each data package ensuring that each is analysed by a different pair of observers, i.e. for each set of 218 spectra we have results from a different pair of collaborators. This was done in order to minimize personal bias in the final classification. Each person then used an interactive tool developed within the VUDS collaboration to carry out the classification.

One problem that may affect our results is the surface brightness limit of our observations that can prevent lower surface brightness emission halos to be seen in the 2D spectra and thus bias our classifications towards a higher fraction of coincident line emitters. Our qualitative separation into the different four classes should thus be interpreted as a classification of \lya\ line emission above our limiting surface brightness. This means that even if most galaxies have extended Ly$\alpha$ halos \citep[e.g.][]{wisotzki2016,wisotzki2018,leclercq2017} there is a distinction between our bright extended emitters and the other class of low surface brightness halos (likely classified as coincident emitters in our study), with different underlying physical conditions likely being responsible for the observed difference in the \lya\ morphology.
\citet{wisotzki2016} found that when restricted to emission matching the spatial extent of the continuum (which is what effectively happens in our 1D extracted spectra), lower equivalent width galaxies have most of their flux in the surrounding halos. If that is the case, and combined with the UV-Ly$\alpha$ size correlation, then we may expect that the coincident line emitters \add{(being those with the lower values of $W_\mathrm{Ly\alpha}$, see Sect. \ref{sssection:lineEWS}) galaxies that do, in fact, include more extended Ly$\alpha$ halos} but at a surface brightness lower than our detection limit. This would mean that we are observing only the core component of Ly$\alpha$ emission which matches roughly the continuum surface brightness profile \citep[e.g.][]{leclercq2017}. \add{We will discuss the effect of these considerations further in Sect. \ref{sec:discussion}.}

\subsection{Classifier agreement}

\begin{figure}
\centering
\includegraphics[width=\linewidth]{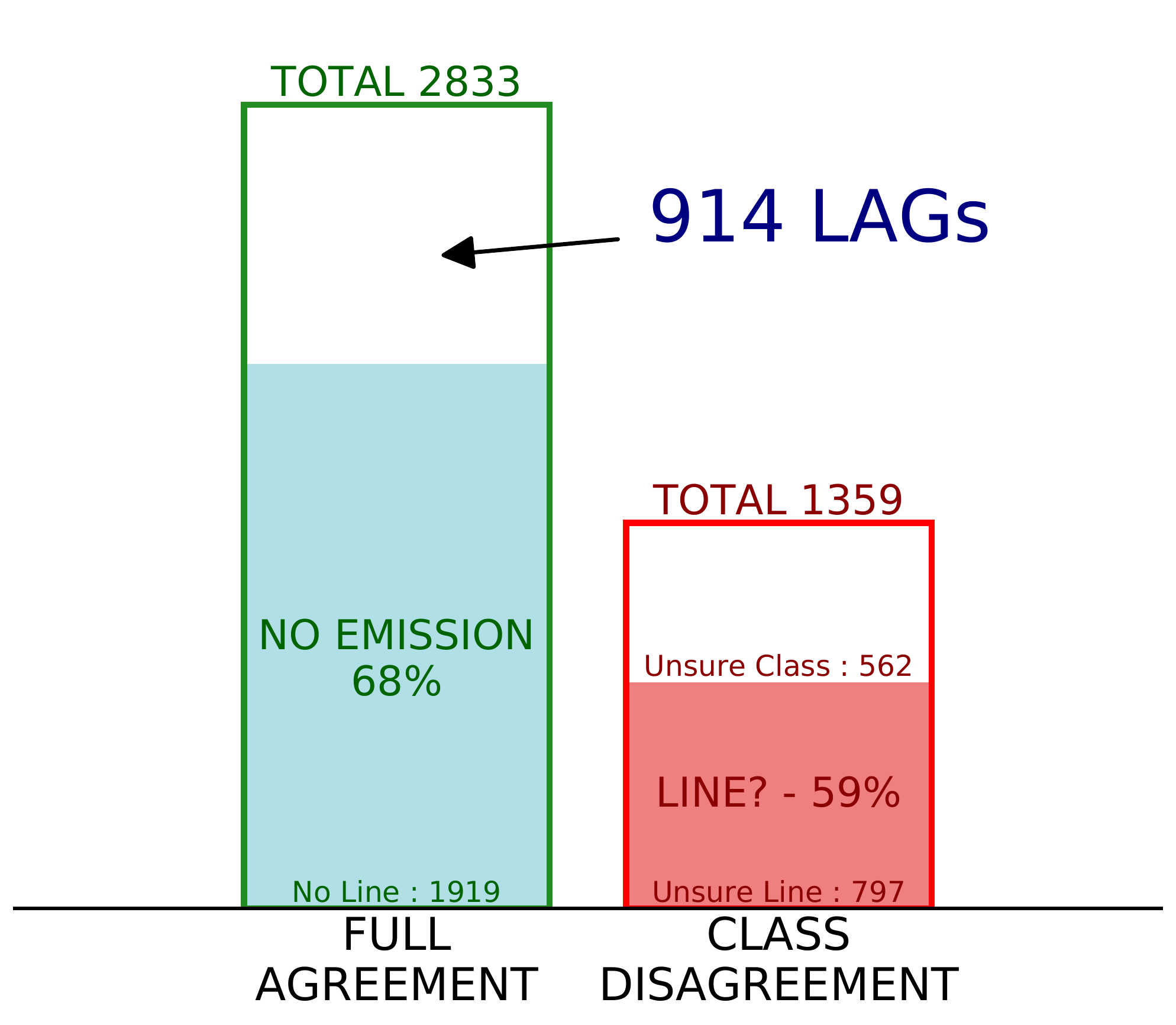}
\caption{Summary of the agreement between collaborators. In the green bar we have the number of galaxies on which both classifiers agreed (divided in emitting and non-emitting galaxies). In the red bar we have the number of galaxies on which the classifiers disagree. {The dark red portion of the bar indicates the fraction of galaxies that one classifier chose an emission line class and the other chose \emph{no emission}, most of which (58\%) are a combination of \emph{faint} and \emph{no emission}}. The white fraction of the bar indicates cases where both collaborators agree on the existence of a line, but disagree on the line classification. {Restricting our sample to the cases where we have secure classifications we find a total of 914 LAGs.}}
\label{fig:class_agreement}
\end{figure}

After the classification was done, we checked for the overall agreement between each pair of classifications for our entire sample. We have classifications for a total of 4192 galaxies{, of which 84 have duplicate observations and thus 4 classifications (2 per observation). The summary of our results is shown} in Fig \ref{fig:class_agreement}. We note that we have agreement in the classification for 2833 ($\sim68\%$) galaxies of which 914 have Ly$\alpha$ in emission and 1919 have no line in emission. For the other 1359 galaxies, there is a disagreement for 797 ($\sim$59\%) on whether there is a line in  emission or not and in most of the cases that happens when a team member chooses the Faint class and the other team member chooses no class. There are also a total of 562 galaxies for which {there is an agreement on the existence of the emission line but there is a disagreement on the classification of the line.} In the end, we have a sample of 914 galaxies with Ly$\alpha$ in emission (LAGs) and for which we a have a {high degree of certitude in the visual classification since both observers (or three out of four in duplicate objects) independently agreed on the same classification. We investigated if there was any redshift trend for galaxies for which we have no secure classification, and we find a constant fraction across the redshift range studied here.}

This classification based on the qualitative presence of Ly$\alpha$ in emission, rather than using the flux or equivalent width of the line, {is expected to take into account the variety of possible configurations in the 2D spectrograms}. The fraction of $\sim$32\% [914/2833] of LAGs in the securely classified sample can be broadly compared to the fraction of LAEs in the same VUDS sample, as reported by \citet{cassata2015}, which varies from 10 to 25\% for bright LAEs with $W_\mathrm{Ly\alpha}\leq-25$\AA\ from $z\sim2$ to $z\sim5$. By definition of the classification performed here, LAGs include all LAEs but also include emitters with $W_\mathrm{Ly\alpha}>-25$\AA, hence the larger fraction of LAGs than LAEs. This is discussed further in the next sections.

\section{Ly$\alpha$ shapes}\label{sec:lya_shapes}

\subsection{Statistics of \lya classifications}

Considering the 914 LAGs with secure classifications for the line emission we find that about half of our sample ($\approx45\%$) shows \add{spatially coincident} Ly$\alpha$ in emission \add{that is centred and within the 2D extend of the UV continuum.} We find that {about one fourth} ($\approx24\%$) of the LAG sample has extended Ly$\alpha$ emission, $\approx13\%$ has an emission line and the UV continuum is not seen, $\approx11\%$ shows an emission \add{that has a projected spatial offset with respect to the center of the 2D UV continuum}, and finally $\approx 7\%$ is too faint to be classified. The number statistics of each sub-sample that we discuss in the following paragraphs as a function of redshift, stellar mass and line properties is summarized in Tab. \ref{tab:numbers}.

\begin{figure}
\centering
\includegraphics[width=\linewidth]{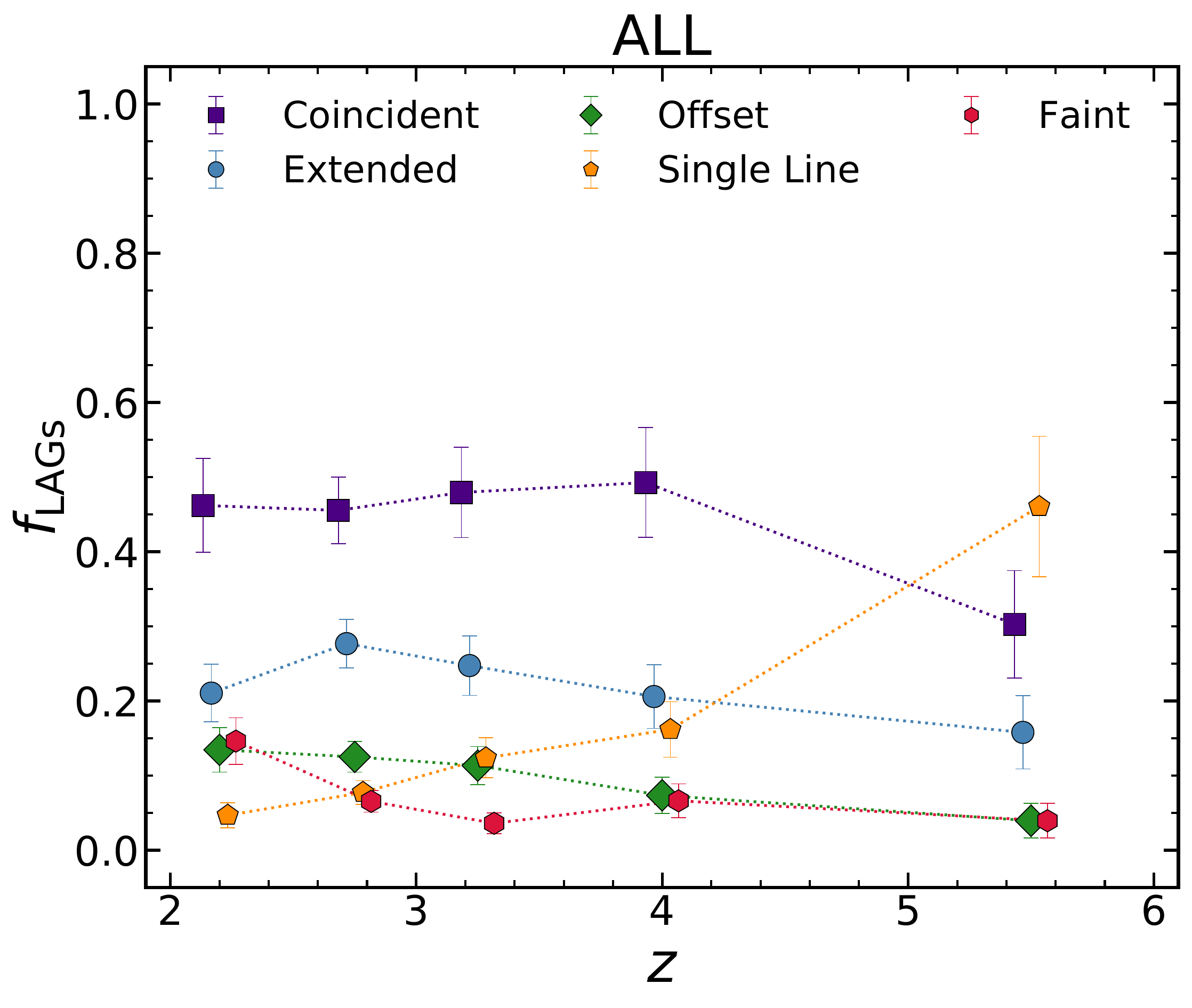}
\caption{The {redshift evolution} of the class fraction of the Ly$\alpha$ line morphology considering the 914 LAGs identified during the classification. Error bars are determined from Poisson statistics. Points are shifted horizontally for visualization purposes and are centred with respect to their redshift bins.}
\label{fig:class_redshift_all}
\end{figure}

\begin{figure*}
\centering
\includegraphics[width=\linewidth]{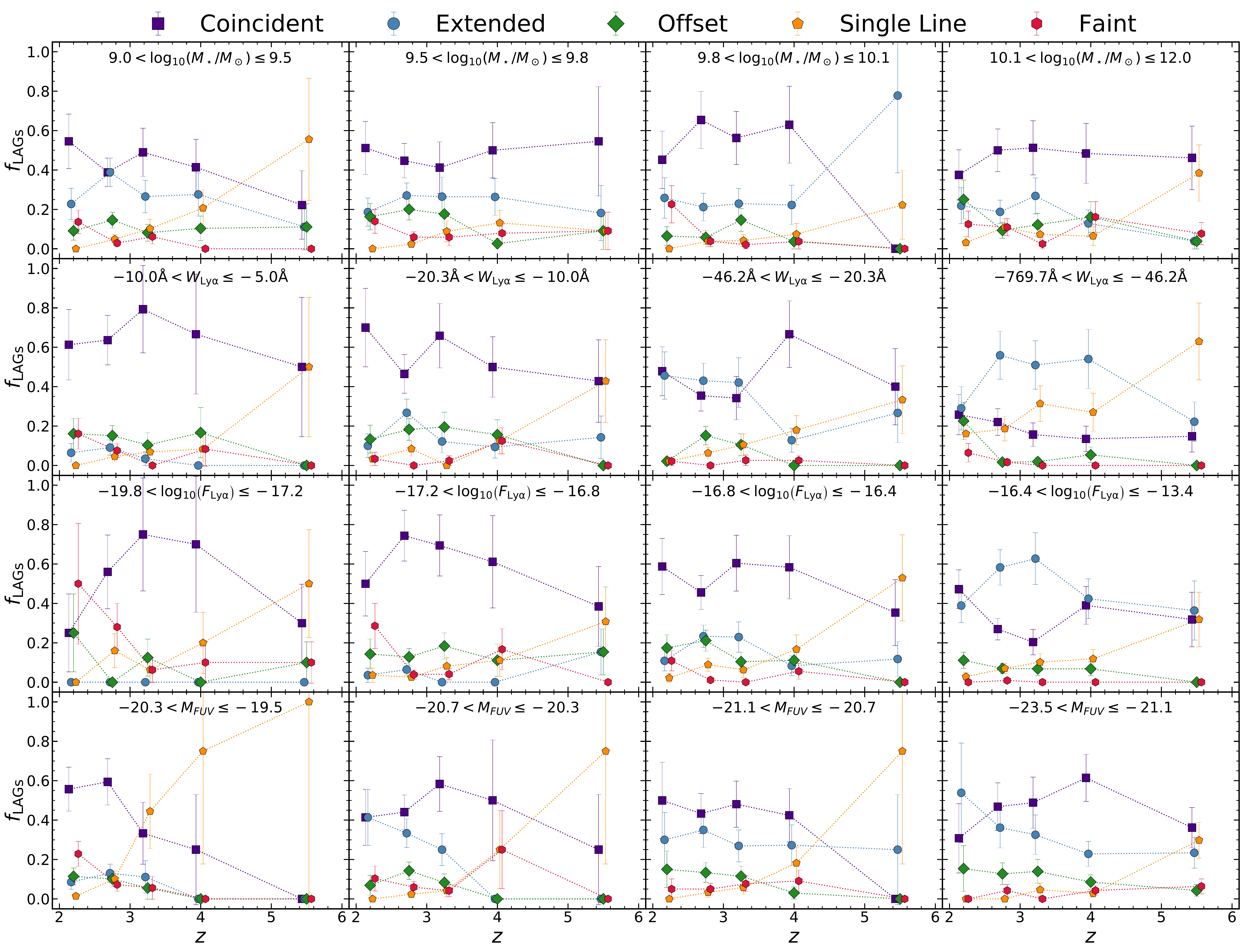}
\caption{The redshift trend of the fraction of each of the main classes of Ly$\alpha$ line morphology, in stellar mass bins (1st row), in line equivalent width bins (2nd row), in line flux bins (3rd row) and in UV absolute magnitude bins (4th row). Error bars are determined from Poisson statistics. Points are shifted horizontally for visualization purposes and are centred with respect to their redshift bins.}
\label{fig:class_redshift_subsets}
\end{figure*}

In Fig. \ref{fig:class_redshift_all} we show how these class fractions vary with redshift. We find that the class with coincident continuum and Ly$\alpha$ emission dominates at $z<4.5$, with fractions always above $40\%$. At $z>4.5$ the most common cases are single line emitters. These show a strong increase in their fraction at high redshift, rising from $\sim5\%$ at $z\sim2.25$ to $47\%$ at $z\sim5-6$. This is caused by the increasing difficulty of detecting continuum in higher redshift galaxies, which naturally increases the number of single line spectra towards higher redshifts \add{see  e.g. dominance of single line emitter in UV faint galaxies in Fig. \ref{fig:class_redshift_subsets}}. The fraction of extended emission is the second most frequent case at $z<4.5$ and it shows an almost constant fraction around $\sim(20-30)\%$ at any redshift. Faint and offset line emitters show a decrease in their fractions as a function of redshift and are never above $\sim 15\%$.

We divide further our sample in subsets based on their stellar-mass, line equivalent width and line flux in Fig. \ref{fig:class_redshift_subsets}. For each property we divide the sample in 4 sub-samples containing approximately 25\% of the sample each (see Table \ref{tab:numbers}).  In terms of stellar mass, coincident Ly$\alpha$ emitters dominate at every stellar mass at redshifts $2<z<4.5$. At higher redshifts ($4.5<z<6$) {single line, coincident, and extended emitters are the dominant population at different stellar masses.}

In terms of line equivalent width, we can see an overall trend that higher equivalent width galaxies ($W_\mathrm{Ly\alpha}\lesssim-45$\AA) are the most common among extended and single line emitters, while coincident line emitters dominate at lower equivalent widths ($W_\mathrm{Ly\alpha}\gtrsim-20$\AA). At intermediate values of $W_\mathrm{Ly\alpha}$, {the dominant populations are coincident and extended emitters} at lower redshifts ($2<z<3.5$) and single line, coincident and extended emitters at $4.5<z<6.0$.

In terms of Ly$\alpha$ flux, \add{we find that at the faint end there is a large dispersion in classifications, with galaxies classified as having faint lines being significant (at $z<2.5$) as expected from its definition, and coincident galaxies are more common at $2.5<z<4.5$. This is likely caused by the evolution of the continuum, with faint lines over a faint continuum being easier to classify as coincident due to the reduced contrast with the UV flux.} At intermediate fluxes we find coincident line emitters to be clearly the dominant population at redshifts $2<z<4.5$. At high Ly$\alpha$ fluxes, we find extended emitters to be the most common population at $2.5<z<4.5$, and coincident line emitters the second most common occurrence. 

Finally, when dividing the sample in \add{FUV absolute magnitude bins} we find that UV-faint galaxies are dominated by single line emitters at $z\gtrsim3$ and coincident line emitters at $z\lesssim3$ which is consistent with effects of cosmological surface brightness dimming preventing us from seeing the UV continuum at increasingly higher redshifts \add{(see discussion in Sect. \ref{ssection:class_proc}, and the trend for single line emitters in Fig. \ref{fig:class_redshift_all})}. For UV-bright galaxies ($-23.5<M_\mathrm{FUV}<-21.1$ mag) we find extended line emitters to be the most common at $2.5<z<4.5$ and coincident line emitters to be the second most common population at similar redshifts. At the highest redshifts ($4.5<z<6$) {we find coincident, extended and single line emitters to have similar population fractions in our sample.}

Galaxies with offsets in line emission are a sub-dominant population ($\lesssim15$\% of the sample) across \add{any studied property.} Nevertheless, we do find them to be more common at low redshifts. They appear to be slightly more common in UV-bright galaxies, but are ubiquitous at any stellar masses and Ly$\alpha$ derived properties. 

\subsection{Measuring spatial offsets from slit data}\label{ssection:offsets}

\begin{figure}
\centering
\includegraphics[width=\linewidth]{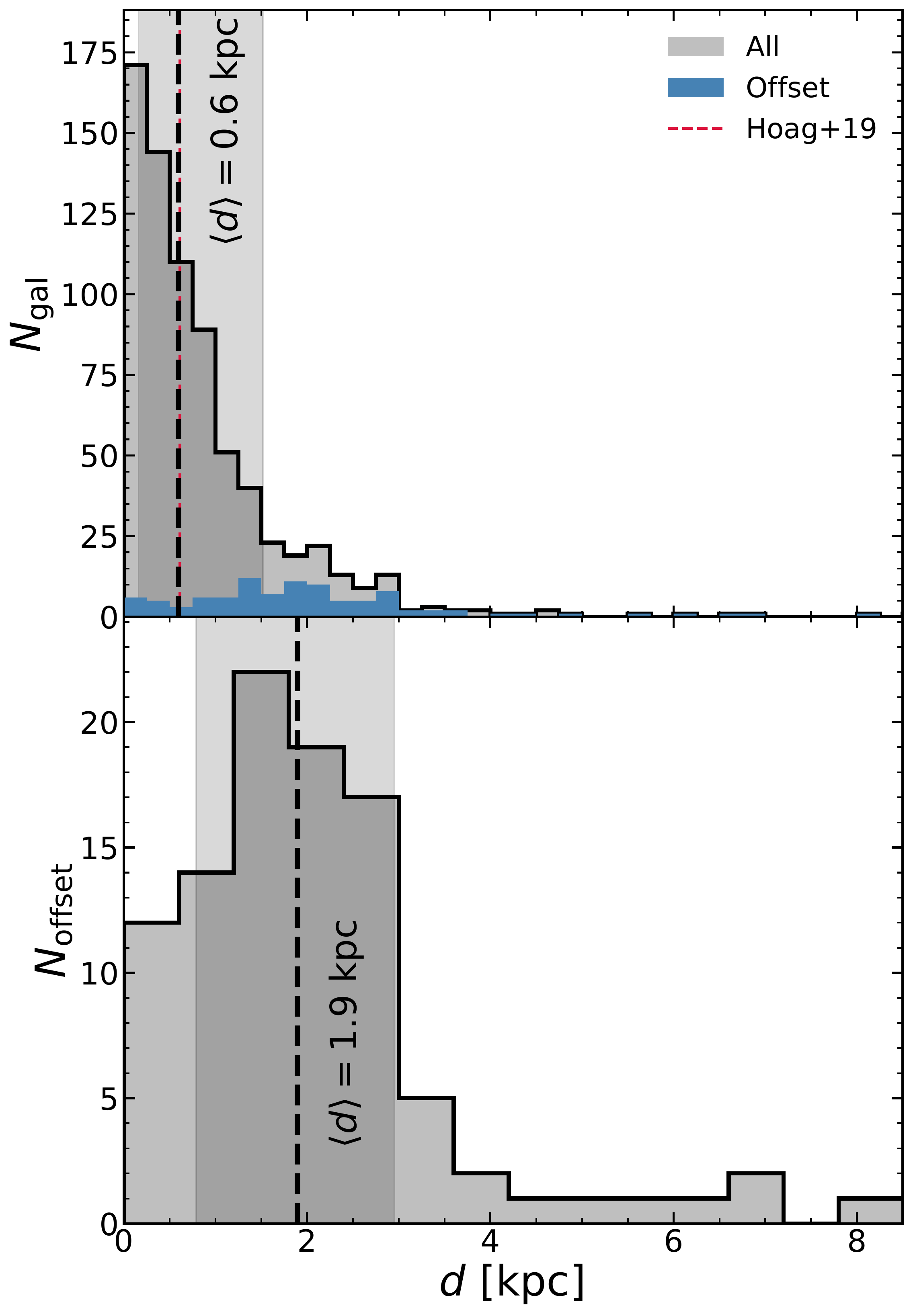}
\caption{Top: the measured absolute distance between continuum and Ly$\alpha$ peaks for all LAGs classified as coincident, extended or offset emitters, with offset emitters highlighted in blue. The vertical line shows the median value of the distribution and the shaded region delimits the 16th and 84th percentiles of the distribution. We show the value derived for 1D offsets measured by \citet{hoag2019} as a red dashed line. Bottom: Same as top panel, but considering only galaxies classified as offset emitters.}
\label{fig:offset_dist}
\end{figure}

\add{
We estimate the offset, $d$, between the Ly$\alpha$ emission and the UV continuum centres by collapsing the 2D spectra along the wavelength direction in two windows: 1212-1225\AA\ for Ly$\alpha$ and 1240-1300\AA\ for the continuum. The choice of wavelength interval for the extraction of the Ly$\alpha$ spatial profile is compromise between maximizing the signal from the Ly$\alpha$ line and minimizing the contamination form UV continuum emission. Before extracting the profiles, we masked all pixels that are close to prominent sky lines in the 2D spectra to avoid contamination. We then fit a Gaussian profile to the emission along the spatial direction and compute the offset distance as the absolute difference between the peaks of each Gaussian. 

Considering our sample of offset emitters we find a median absolute physical separation of $1.9 \pm 0.2$ kpc (with 68\% of the sample being between 0.8 kpc and 3.0 kpc), see Fig. \ref{fig:offset_dist}. There is a recent study measuring the offsets from 2D spectra \citep[see][]{hoag2019}, but there is no fair comparison that can be directly made to our sample of offset emitters due to selection bias and method disparities. To compare the two samples fairly, we computed the offset in Ly$\alpha$ for all galaxies securely classified with a line (includes coincident, extended and offset classes) and we find a median offset of $0.60\pm0.05$ kpc, which is remarkably close to what is reported from the 1D separation of $0.61\pm0.05$ kpc from the \citet[][priv. communication]{hoag2019} sample (see Fig. \ref{fig:offset_dist}). When considering all secure line emitters, we also find a small evolution with redshift, with the median offset rising from $0.51\pm0.06$ kpc at $3<z\lesssim5.5$ to $0.64\pm0.07$ kpc at $2<z<3$. These results are also consistent with the values derived by \citet{lemaux2020}  for a sample of galaxies at higher redshifts ($5<z<7$) and point to a non-significant evolution in the median offset since the epoch of re-ionization towards the peak of cosmic star-formation.

We find that the visual classification of offset emitters is complete for offsets larger than 1.5 pixel ($\sim$3 kpc for the median redshift of the sample), and people classified objects with offsets down to separations of 0.5 pixels ($\sim$1 kpc for the median redshift of the sample, albeit with higher incompleteness). This will mean that the observed distribution of offset distances (see Fig. \ref{fig:offset_dist} - bottom panel) will capture most large offsets but largely be incomplete for separations smaller than 3 kpc. Nevertheless, the existence of a tail extending towards larger distances is indicative of the possible existence of mergers among the offset emitters, combined with bright Ly$\alpha$ clumps that are found at smaller offsets. We note that the incompleteness at low offset distances would only exacerbate the existence of the two scenarios, with a large concentration of small offsets for bright in-situ clumps combined with a larger tail for on-going merger events. This is consistent with the scenario described by \citet{ribeiro2017} who studied the occurrence of bright UV clumps for the same parent sample. 
}

\begin{figure*}
\centering
\includegraphics[width=\linewidth]{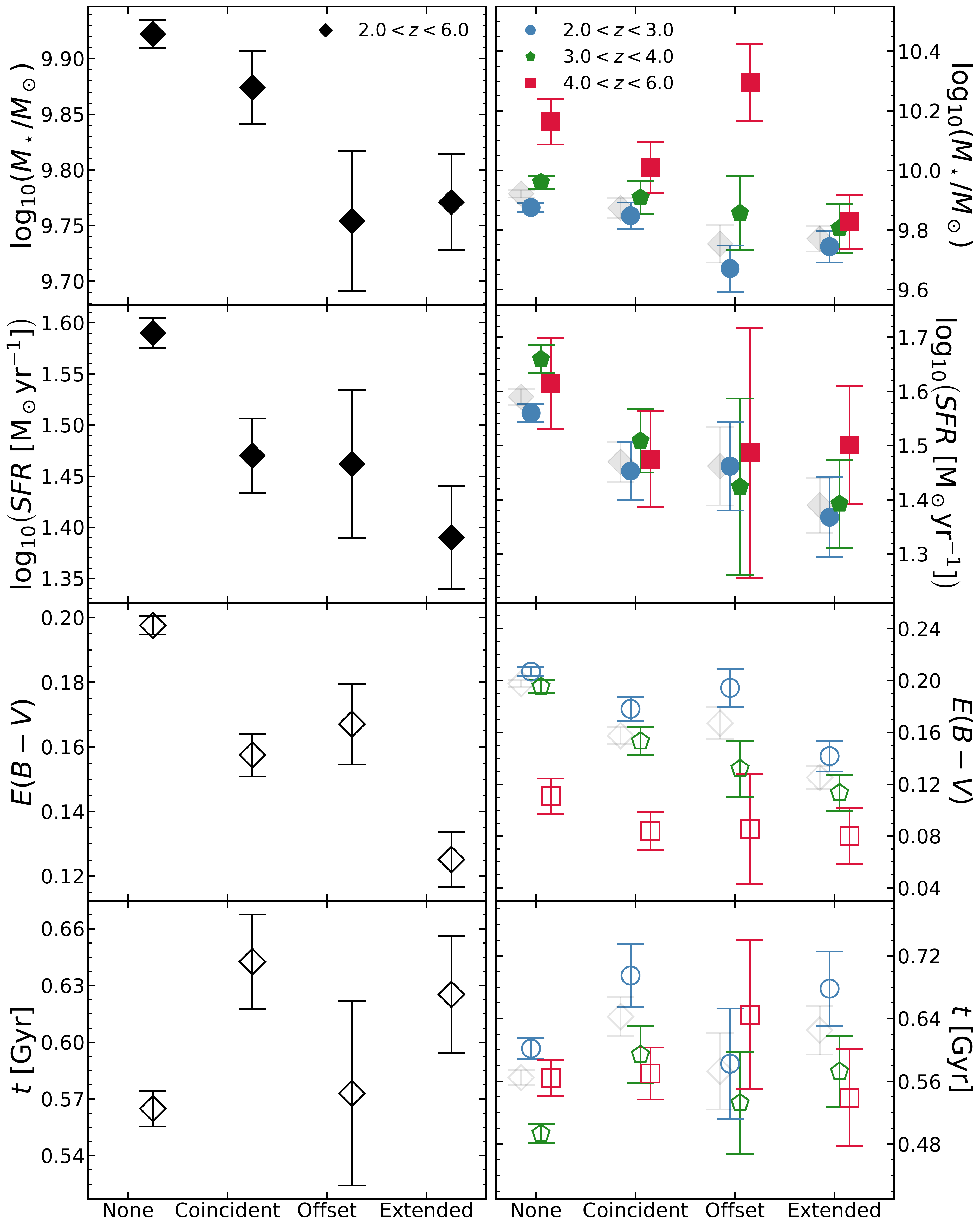}
\caption{Summary of median/average properties for different classes of the 914 LAGs with secure classifications. From top to bottom we show stellar mass, star formation rate, dust attenuation and mean stellar age. On the left panels we show individual redshift bins, and in the right panels we show an highlight of the sample across the full redshift range. \add{The data from the left panels is plotted again in the right panels (but note the change in the y-axis scale) for a straight comparison to individual redshift slices.} Filled symbols represent the median of the population and empty symbols represent the mean of the population (see Sect. \ref{sec:sed_physics}).}
\label{fig:class_physics_redshift}
\end{figure*}

\section{Physical properties of Ly$\alpha$ line morphology classes}\label{sec:lya_physics}

\subsection{SED derived parameters}\label{sec:sed_physics}

We show in Fig. \ref{fig:class_physics_redshift} the median values for stellar masses, SFRs, dust extinction and stellar age for the four main classes defined in Sect. \ref{sec:lya_classification}. {We first discuss the trends for the entire sample (left panels) and then focus on the results tracing the evolution in redshift (right panels).}

When looking at the stellar masses it is clear that galaxies with no \lya\ emission and coincident emission {tend to be} slightly more massive (median of $\log_{10}\left( M_\star/M_\odot \right) \sim 9.90$) than galaxies with offset or extended emission (median of $\log_{10}\left( M_\star/M_\odot \right) \sim 9.75$). In terms of SFR, there is an interesting trend {showing that the most star-forming galaxies have no \lya\ emission at any redshift}. Coincident emission line galaxies are the second most star-forming population followed by offset and extended emission lines. {The latter is the population with the lowest star-formation rate. This might be correlated with the average $W_\mathrm{Ly\alpha}$ of each class, as other studies find an anti-correlation between SFR and \lya\ equivalent width \citep[e.g.][]{kornei2010,hathi2016,pahl2020}. We note, however that among the classes of \lya\ emitters we find average SFRs consistent among each class which indicates that non-emitters have higher SFRs than \lya\ emitters \citep[as found by e.g. ][]{marchi2019}.}

Since the values for dust extinction and stellar age are derived from a grid of models, the median value is fixed at one of these grid values. Thus, we opt for the mean value for these two quantities in our interpretation. In terms of dust extinction it is clear that extended emission line galaxies are the ones with the least amount of dust. Then coincident and offset line galaxies are more dusty and the population with larger amounts of dust extinction is when no line emission is observed. In terms of stellar age we find that coincident line galaxies are likely the older population (but consistent with extended line emitters and to some extent offset emitters) and galaxies with no emission are on average the younger population. \add{However, the differences of stellar ages among each class are not statistically significant and the more likely scenario is that all of the classes are comprised of relatively young galaxies with a considerable age spread}.

{In terms of the redshift evolution of the observed trends we show in the left panels of Fig. \ref{fig:class_physics_redshift} the same quantities computed in three redshift bins}. In terms of stellar mass there is a trend for higher stellar masses at higher redshifts for the four classes, it is likely the result of the VUDS sample selection function including a limiting magnitude at $i_{AB}\leq 25$. The relative ratio of stellar masses between coincident and extended line emitters remains roughly constant with the latter being the least massive population. When looking at the median SFRs, the trends are broadly the same as found for the full redshift range with extended line emitters being the least star-forming population whereas galaxies with no line emission are the most star-forming.

We find again that at $z<4$ the extended line emitters have the least dust content and that dust content increases gradually to the offset, coincident and no line populations. We find no statistically significant differences in the mean age for each class. At $z>4$ the trends are not as strong but we still find, despite the error bars, that the extended line emitters have the lowest amount of dust. Interestingly, the dust attenuation in galaxies rapidly declines for $z>4$ for each class.

\subsection{Ly$\alpha$ escape from galaxies}

\begin{figure}
\centering
\includegraphics[width=\linewidth]{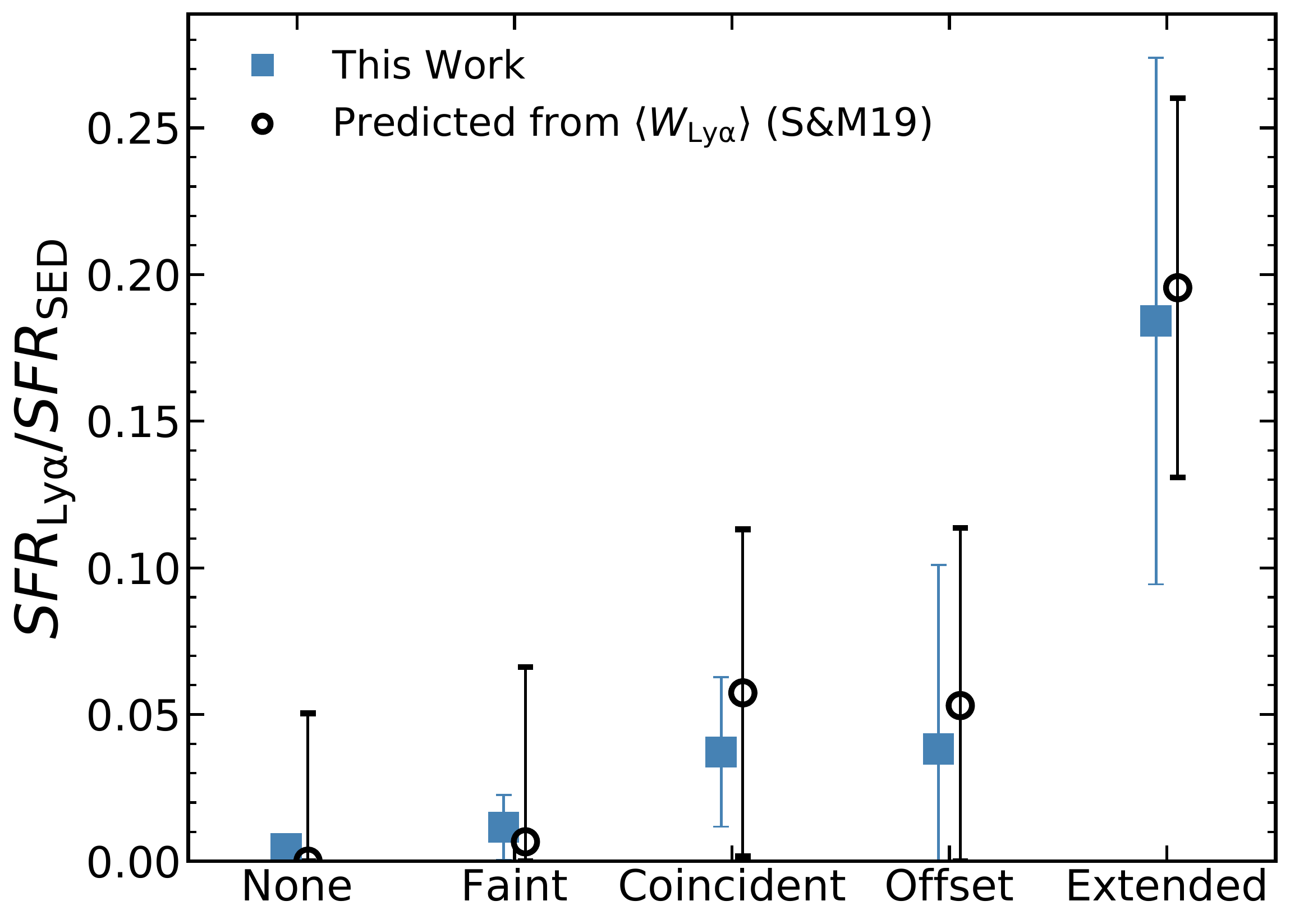}
\caption{Median ratio of SFR derived from Ly$\alpha$ and from SED fitting for each class with detectable UV continuum (blue squares). We show as black circles the predicted ratio based on the median equivalent width of the Ly$\alpha$ emission in each sample, as parametrised by \citet{sobral2019}. The uncertainty on the prediction combines the error on the median and the error on the parametrisation.}
\label{fig:LyaUvRatio}
\end{figure}

We compute as a proxy of the Ly$\alpha$ escape fraction the ratio of the SFR derived from SED fitting and those derived from the Ly$\alpha$ luminosity. The Ly$\alpha$ derived SFR is estimated assuming a case B recombination, in conditions that the intrinsic ratio between Ly$\alpha$ and H$\alpha$ is 8.7 \citep[e.g.][]{lemaux2009,hayes2011,cassata2015,matthee2016}. We then use the \citet{kennicutt1998} relation to derive $SFR_\mathrm{Ly\alpha} = 5.985\times10^{-43} L_\mathrm{Ly\alpha}\ \mathrm{M_\odot yr^{-1}}$ \add{(divided by a factor of 0.63 to correct for a \citealt{chabrier2003} IMF, see e.g. \citealt{madau2014}).}

We show in Fig. \ref{fig:LyaUvRatio} the resulting median ratio for each class that has a detectable UV continuum. \add{As expected, galaxies classified with no visible emission have this ratio equal to zero.  We find a $SFR_\mathrm{Ly\alpha}/SFR_\mathrm{SED}$ ratio of $4\pm2$\% for coincident and $4_{-4}^{+6}$\%  for offset emitters. The extended class of emitters has a higher ratio of $19\pm9$\% . WE note that most of our emitters (excluding the Extended class) have a similar escape fraction to that computed for a field sample at $z\sim4.5$ by \citet{lemaux2018}.} The latter higher observed escape fraction for extended line emitters is {consistent with the higher equivalent width of this class of emitters (see also Sect. \ref{sssection:lineEWS}), as studies found these quantities to correlate with each other \citep[e.g.][]{sobral2017,verhamme2017,sobral2019,cassata2020}.}

\subsection{Composite spectra}\label{sec:lya_stacks}

\begin{figure*}
\centering
\includegraphics[width=\linewidth]{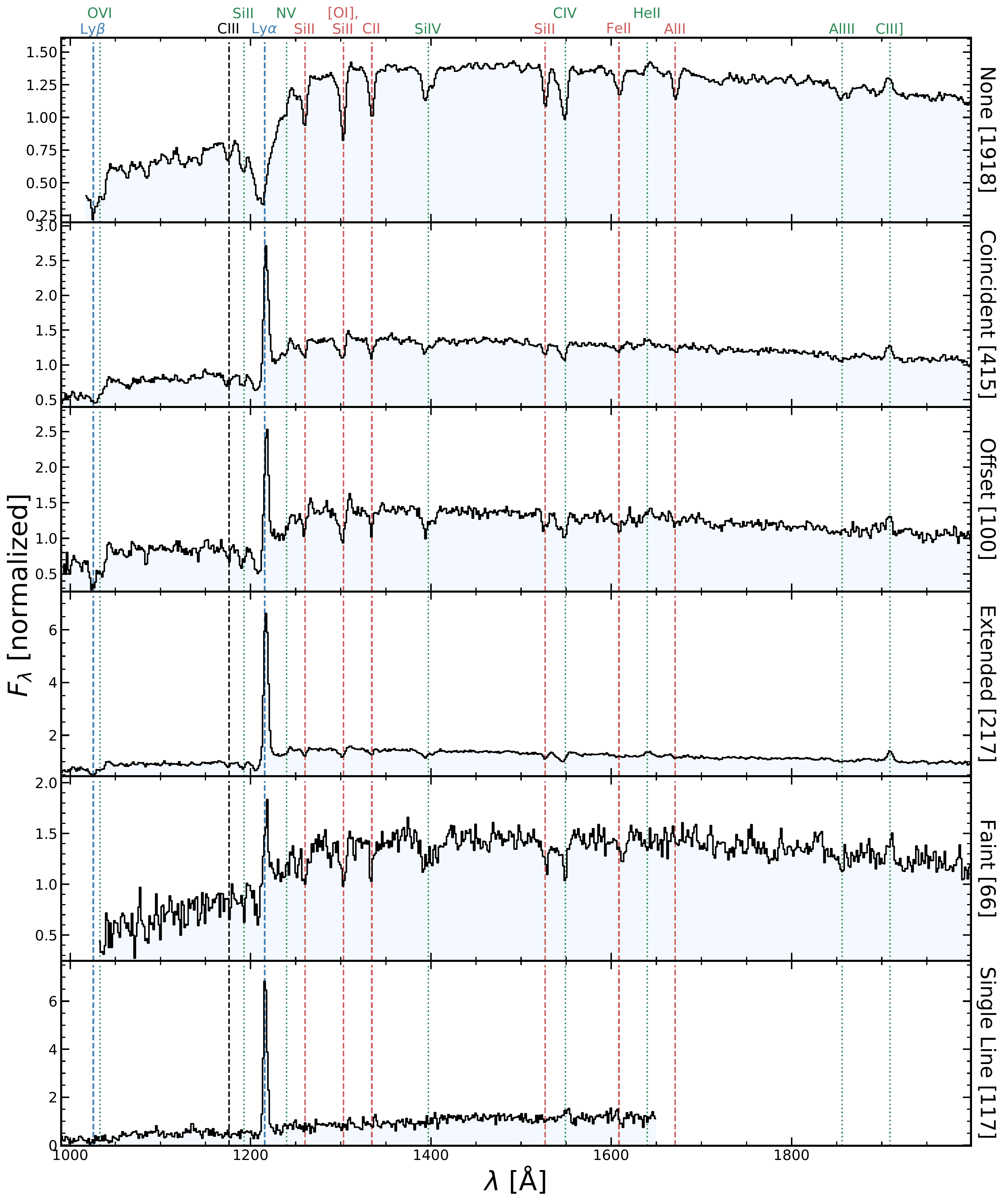}
\caption{Composite spectra of galaxies grouped by their Ly$\alpha$ line morphology class. The number of spectra in each stack is shown next to the class label. In each panel, the vertical dotted lines mark the position of observed spectral lines. {Lines from the Lyman series are marked in blue, lower ionization inter-stellar absorption lines studied in this paper are shown in red, photospheric absorption lines are shown in black and the remainder are shown in green.}}
\label{fig:class_stacks}
\end{figure*}

\begin{figure*}
\centering
\includegraphics[width=\linewidth]{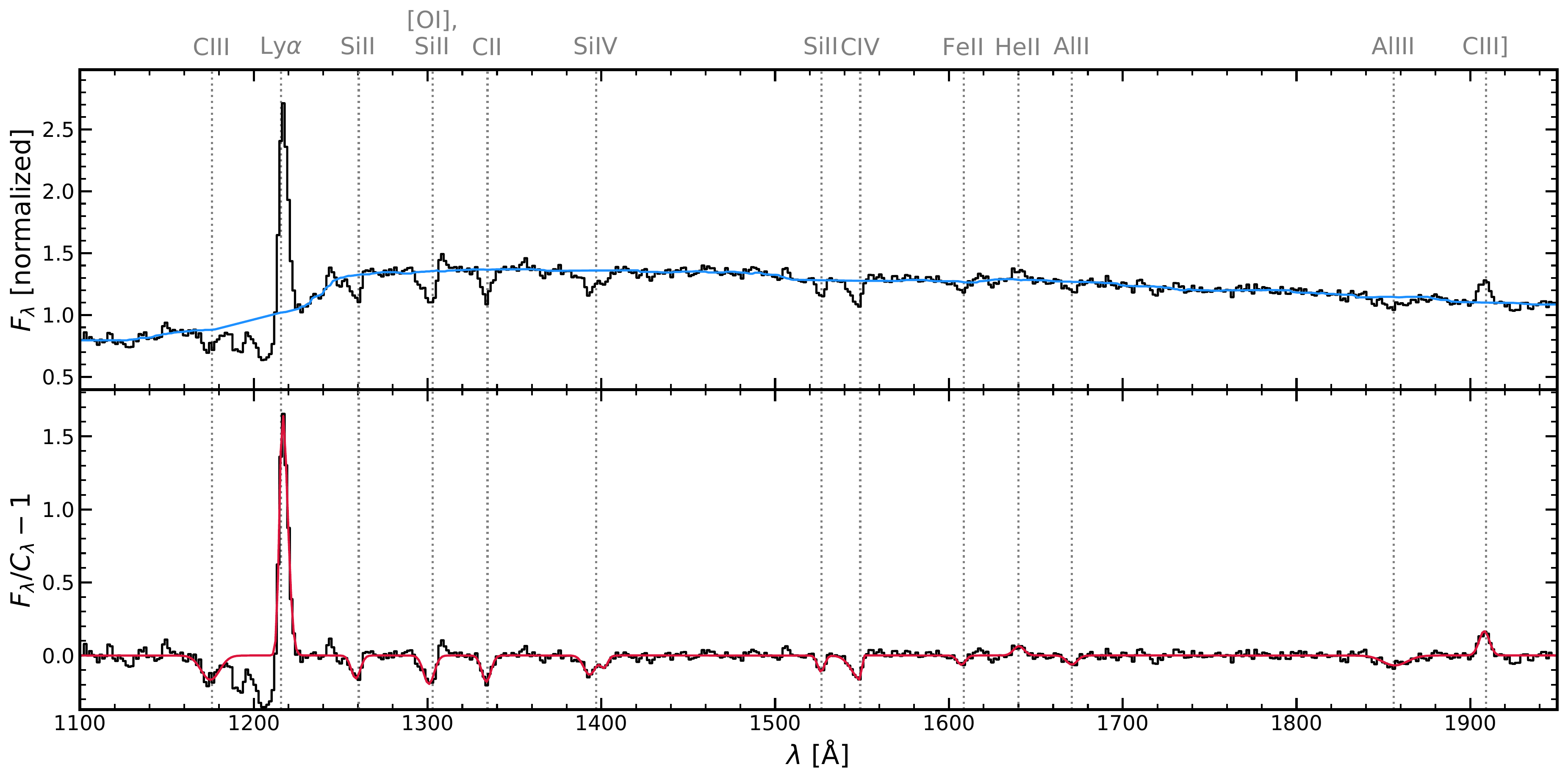}
\caption{Example of the fitting routine on each of the lines marked on the plot {for the stack of coincident line emitters}. On the top panel we show our derived continuum level (blue solid line) from smoothing a masked version of the original spectrum (black solid line). On the bottom panel we show the continuum-normalized spectra (black solid line) and the combination of all the {individual models used to fit each line} (red solid line). The vertical dotted lines mark the position of all spectral lines for which a model was fit. {We show the composite spectra for each individual class in Fig. \ref{fig:class_stacks}.}}
\label{fig:class_stacks_fit}
\end{figure*}

To assess the average spectral properties we have produced composite spectra for each class defined in  Sect. \ref{sec:lya_classification}. 

We combined all of the spectra by linearly interpolating individual spectra (normalized by the integrated flux over 1450-1500\AA) onto a common grid of 1.5\AA /pixel elements and then taking the median flux at each wavelength. We show  the composite spectra in Fig. \ref{fig:class_stacks}. {Prior to combining the individual spectra, we refined our redshift estimates by cross-correlation of individual spectra to a master template of the sample \citep[e.g.][]{tonry1979,hewett2010,talia2017}. In summary, we stacked every spectra with a secure classification to create a master template. Then we estimated the continuum using a 61 pixel median filter \citep{hewett2010} and normalized the final template by this continuum. We then restricted our template spectra to the 1270-2050\AA\ wavelength range. \add{We note that this range includes both ISM absorption lines and C{\sc iii}] emission, to maximize the chance of aligning each spectrum. Since these are two different velocity components the re-fined spectrum will be weighted by the number of lines detected. And while we expect some broadening of the lines due to velocity differences of all the lines used in the master template, the centroid of each line should be robust}. We repeated the same continuum estimation and wavelength restriction to individual spectra and clean the spectra with standard sigma-clipping ($\sigma=3$) to remove noise spikes. Since we do not expect strong emissions and/or absorptions in the wavelength range considered, this automatic cleaning should not affect lines in the spectra. Finally, we computed the cross-correlation function, $cc(s)$. We assumed a maximum shift of $\Delta z\pm0.01$ and computed the maximum $s_\mathrm{max}$ of $cc(s)$ by fitting a quadratic function around the peak of $cc(s)$ ($ s = s_\mathrm{max} \pm 0.001$). We note that we apply our redshift correction to galaxies where $cc(s_\mathrm{max})>0.25$ to discard noisy cross-correlations  \citep[see e.g.][]{hewett2010}. After correcting each individual spectra, we further refine them with a new master template re-computed from the new refined individual spectra. We repeated this procedure until we converged on a final list of redshifts.}

For all subsequent analysis we use the wavelength regions which have more than 75\% of the spectra contributing to that wavelength bin. {With this constraint we} cover the rest-frame range 1000-2000\AA\ for all of the classes but one. The class of single line emitters is dominated by higher redshift objects and so the common range among different objects is smaller than for the other classes. The fitting procedure is described below and illustrated in Fig. \ref{fig:class_stacks_fit}.

To estimate the continuum level around emission and absorption lines we constructed a continuum function of wavelength by masking all regions around the lines (15\AA\ around simple lines, 30\AA\ around line doublets and the region 1170\AA\ - 1220\AA\ around Ly$\alpha$) and then smoothing the spectrum with a median filter of 25 pixels (corresponding to a kernel size of $\sim38$\AA). We then used the continuum normalized spectra to compute line equivalent widths and velocity offsets from a set of different Gaussian models.

We fit each line with one of three different models, all based on Gaussian shapes. The first model is a single Gaussian function with three free parameters. The second model is the combination of two half-Gaussian functions each with its own width, and thus has four free parameters. {Finally, the third model is a simple combination of two Gaussian functions, with a total of 6 parameters (of which only four are free since we fix the doublet separation and individual component widths to be the same). The half-Gaussian model is used for the Ly$\alpha$ emission line. The double Gaussian model is used for the C{\sc iii}], S{\sc iv} and Al{\sc iii} doublets.} The other lines are fitted with a simple Gaussian.

{We assume that the systemic velocity of each composite spectra is that of the \ciii\ emission line, which should trace nebular emission in H{\sc ii} regions around star-forming regions \citep[see e.g.][]{talia2017}.} We could potentially use the photospheric absorption C{\sc iii}$\lambda$1176 but we find it to be strongly affected by the Ly$\alpha$ red absorption wing which complicates the measurement of the systemic redshift in this line.

\subsubsection{Line equivalent widths}\label{sssection:lineEWS}

\begin{figure}
\centering
\includegraphics[width=\linewidth]{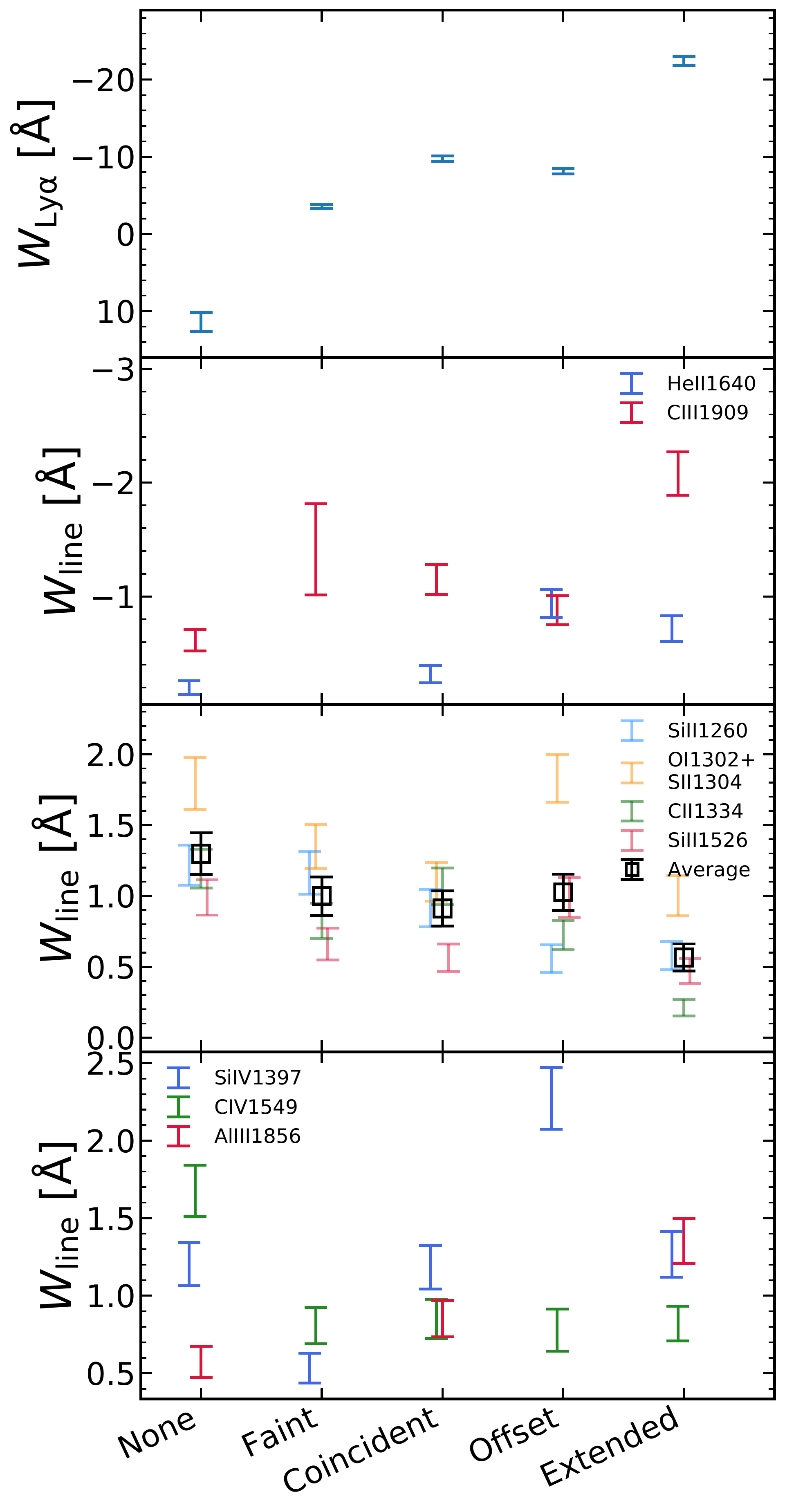}
\caption{Line equivalent widths for observed features in our composite spectrum per morphological Ly$\alpha$ emission class. From top to bottom: Ly$\alpha$, emission lines (He{\sc ii} and C{\sc iii}]), low ionization and high ionization interstellar absorption lines.}
\label{fig:all_eqws}
\end{figure}

We show in Fig. \ref{fig:all_eqws} the line equivalent widths for all of the lines listed in Table \ref{tab:eqwidth} that are present in the composite spectra. {Galaxies with no emission have a significant absorption in Ly$\alpha$ with an equivalent width of $11\pm1$\AA. For faint emitters, we observe Ly$\alpha$ to have low equivalent widths ($-3.5\pm0.3$\AA), as expected from their definition. Single line emitters show the strongest Ly$\alpha$ equivalent width ($-75\pm1$\AA), also expected due to the lack of a strong continuum detection on these sources. Galaxies with coincident and offset emission have equivalent widths of $-9.7\pm0.4$\AA\ and $-8.1\pm0.4$\AA, respectively. Galaxies which show an extended emission with respect to the continuum have $W_\mathrm{Ly\alpha}=-22.4\pm0.6$\AA. {We expect the values for offset and extended emitters to be underestimated by $\sim$20\% and $\sim$8\% (see Sect. \ref{ssec:lya_measures}), yielding values of $W_\mathrm{Ly\alpha}=-9.7$ and $W_\mathrm{Ly\alpha}=-24.2$\AA, respectively.}}

For the other two emission lines seen in our composite spectra (C{\sc iii}] and He{\sc ii}) {we find that extended emitters are those showing strong emission also in these lines (especially in C{\sc iii}). The other classes show similar properties among themselves.} This relation can be likely traced back to a {tentative correlation between equivalent widths in Ly$\alpha$ and C{\sc iii}] \citep[seen on stacks, see also][]{guaita2017,lefevre2017}.}

Low ionization interstellar medium (ISM) absorption lines are on average stronger for galaxies with no Ly$\alpha$ in emission and progressively weaker for faint, offset, coincident and extended line emitters. This trend is inverted with respect to Ly$\alpha$ equivalent width which is likely the explanation for these differences among different classes, as there are several reports on the existence of weaker low ionization ISM absorption lines in stronger Ly$\alpha$ emitters \citep[e.g][]{shapley2003,jones2012,jones2013,leethochawalit2016,pahl2020}. Higher ionization ISM lines show no strong correlation on the class of emitters. 

\subsubsection{Velocity offsets}

\begin{figure}
\centering
\includegraphics[width=\linewidth]{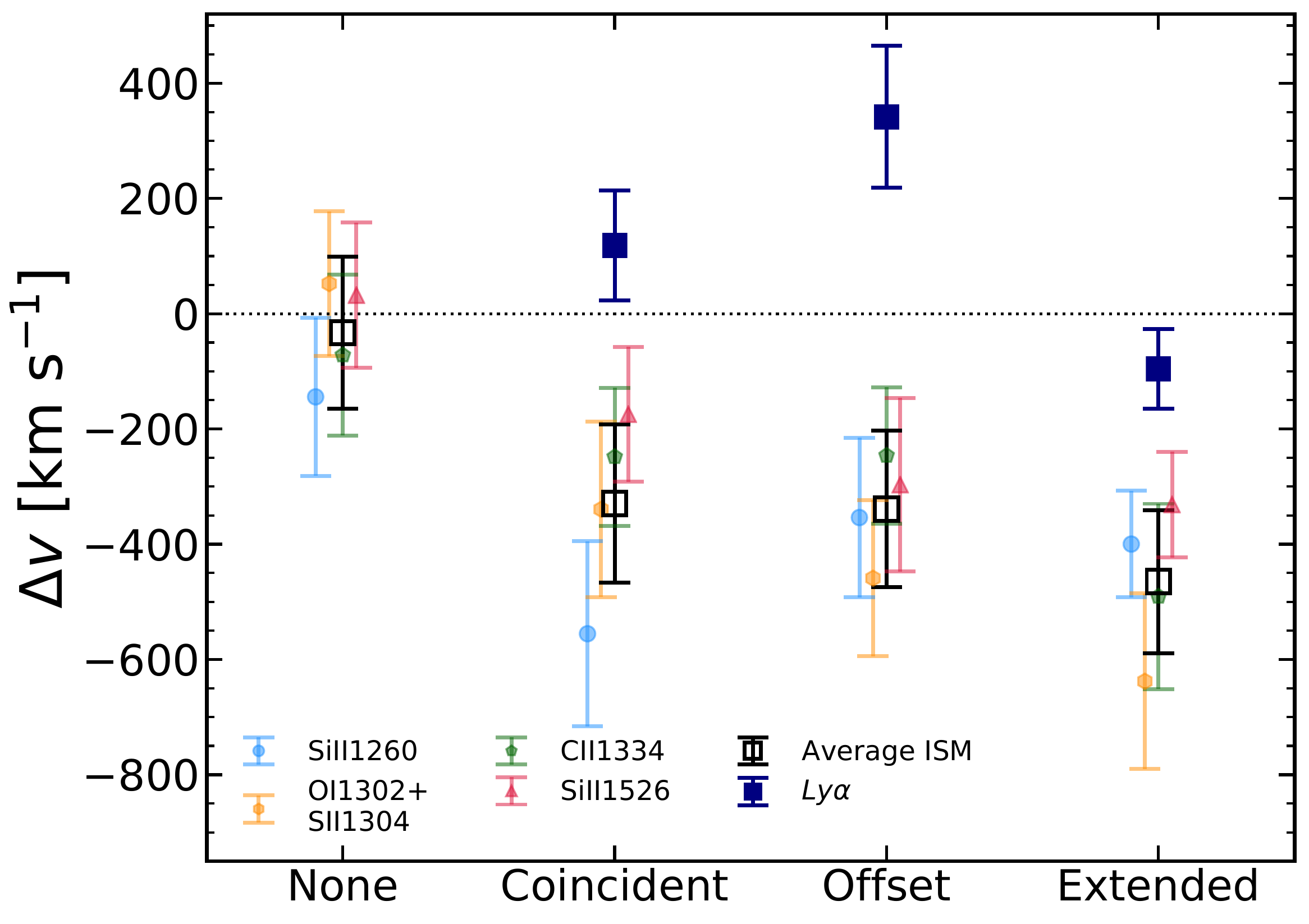}
\caption{Velocity offsets (with respect to C{\sc iii}] emission) for Ly$\alpha$ and low ionization ISM absorption lines in the composite spectra of each morphological Ly$\alpha$ emission class.}
\label{fig:all_voffs}
\end{figure}

In Fig. \ref{fig:all_voffs} we show the velocity offsets of the blue emission peak of Ly$\alpha$ and of low ionization ISM lines  with respect to the \ciii\ emission, assumed here as a tracer of the systemic velocity. We find that galaxies with no \lya\ emission have an average velocity offset of $\sim30\pm130 \mathrm{km\ {s^{-1}}}$ which is roughly consistent with star-forming galaxies at similar redshifts \citep{shapley2003, vanzella2009, talia2012, talia2017}. Galaxies with \lya\ in emission show higher velocity offsets for the low-ionization lines ($\sim300-450\mathrm{km\ {s^{-1}}}$) with non-significant differences among the different classes. The blue-shifted ISM lines with respect to C{\sc iii}], might be indicative. Coincident and extended \lya\ emission galaxies show similar differences in velocity between Ly$\alpha$ and low ionization lines despite having different equivalent widths in Ly$\alpha$, in agreement with other studies \citep[see e.g.][]{shapley2003,jones2013,marchi2019}.

One interesting result is that for galaxies with offset Ly$\alpha$ emission we find a larger difference between \lya\ and low ionization ISM lines velocity offsets. This is mostly driven by a red-shifted \lya\ emission with respect to C{\sc iii}].
We hypothesize that these offset regions that emit Ly$\alpha$ are kinematically decoupled from the galaxy ISM. These regions have {a typical spatial offset of $\sim1.1$kpc (see Sect. \ref{ssection:offsets}) which is similar to the observed median size} for the same galaxies at these redshifts \citep[][\!\! see also Fig. \ref{fig:offset_dist}]{ribeiro2016}. {The combination of these two quantities suggest that we are probably observing} offset emission from a lopsided distribution of HI gas in the galaxy (e.g. infalling material along a filament, accretion of fainter satellite or a bright off-center star-forming clump). {It might also be produced by inhomogeneous dust distribution alone, as we do not find difference in the ISM kinematics relative to the other classes of line emitters. Another possibility is if the observations were produced by gas scattering on the circum-galactic medium, but in this scenario we should expect to see larger average spatial offsets than what we find (as we do observe for individual galaxies, but it is not the dominant process)}.

\subsection{The UV size of Ly$\alpha$ emitting galaxies.}\label{sec:lya_sizes}

\begin{figure}
\centering
\includegraphics[width=\linewidth]{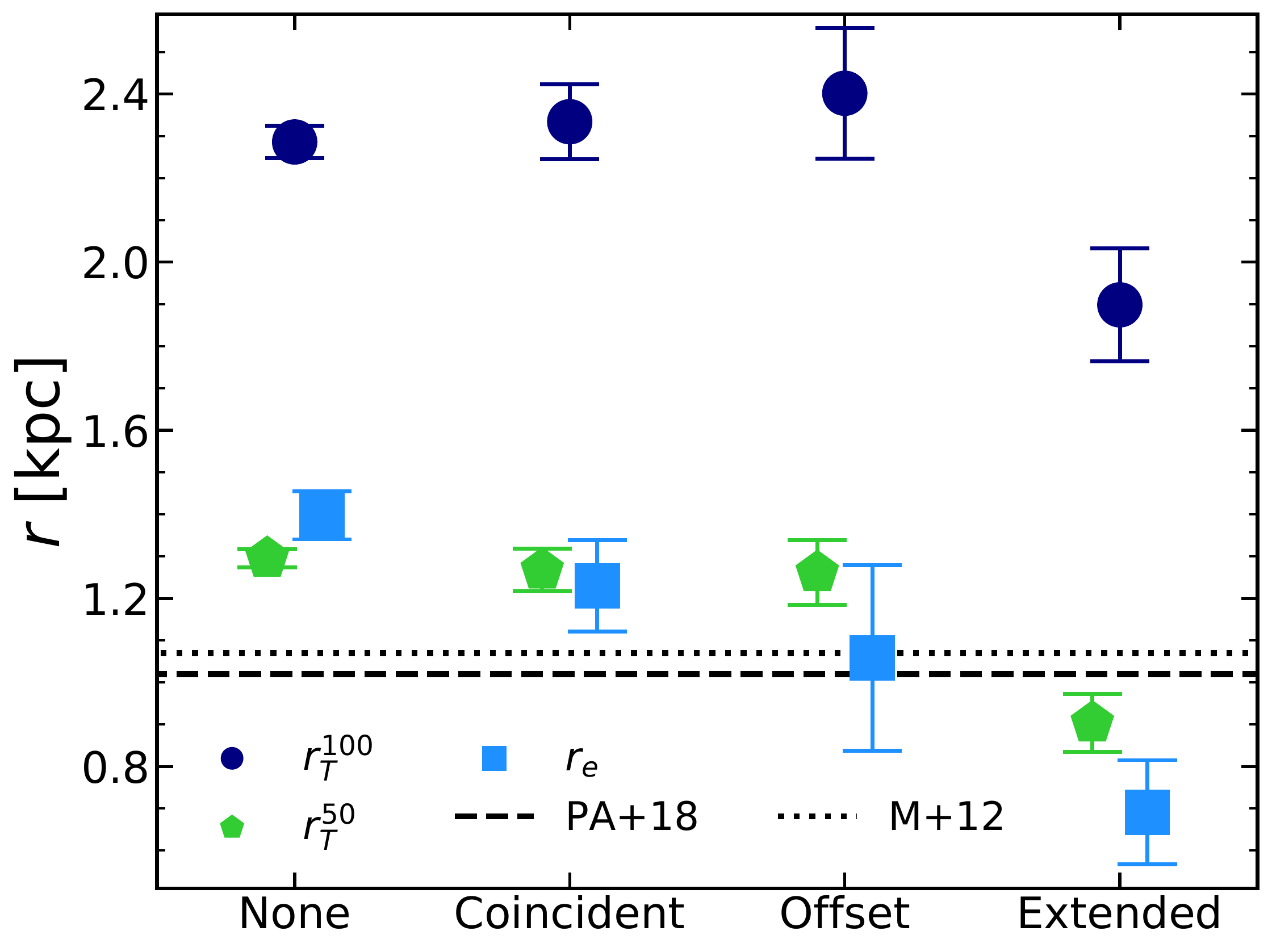}
\caption{Median size of galaxies measured in the rest-frame UV (measured in HST F814W) as a function of classification. We show values for the effective radius (in light blue), total extent (in dark blue) and 50\% light radius (green) for galaxies with stellar mass $\log(M_\star/M_\odot)>9.5$ {(see Sect. \ref{ssection:galSizes})}. In all cases the extended line emitters have smaller UV sizes. We show as dashed/dotted lines the median half-light radius of LAEs derived by \citet{paulino-afonso2017} and \citet{malhotra2012}, respectively.}
\label{fig:class_sizes}
\end{figure}

\begin{figure}
\centering
\includegraphics[width=\linewidth]{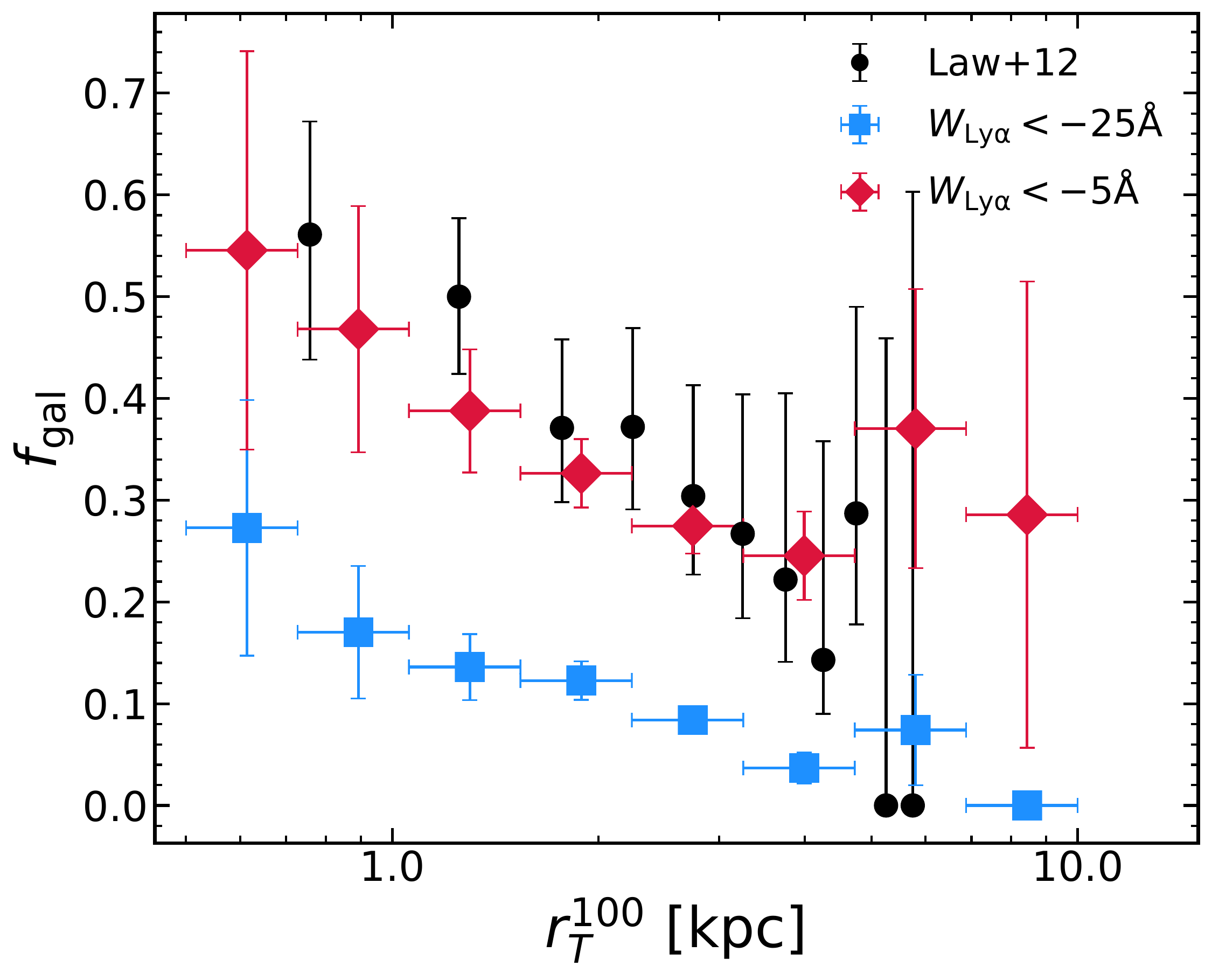}
\caption{Fraction of LAEs as a function of total  rest-frame UV extent (measured in HST F814W) for galaxies with stellar mass $\log(M_\star/M_\odot)>9.5$ {(see Sect. \ref{ssection:galSizes})}. Blue squares and red diamonds show the resulting fraction using two different Ly$\alpha$ equivalent width cuts (-25\AA\ and -5\AA\, respectively). We see on both cases that the occurrence of Ly$\alpha$ in emission is more common in smaller galaxies. All size bins have at least 5 galaxies. We show for comparison the results from \citet{law2012b}, where they report the fraction of galaxies with \emph{visible} Ly$\alpha$ emission.}
\label{fig:lae_fraction_size}
\end{figure}

To compare the UV size of different classes of emitters, we use the size measurements detailed by \citet{ribeiro2016} {and briefly described in  Sect.  \ref{ssection:galSizes}. We show in Fig. \ref{fig:class_sizes} the median sizes for each of the classes of Ly$\alpha$ emission}. Interestingly we do find that the  extended line emitters have smaller UV extent than the other classes. The non-emitters are the largest in terms of the 50\% light radius (both $r_e$ and $r_T^{50}$) {but have similar sizes} as the coincident and offset line emitters. We further explore the dependence of Ly$\alpha$ emission on the UV rest-frame size in Fig. \ref{fig:lae_fraction_size} where we show the fraction of LAEs, defined by using two Ly$\alpha$ equivalent width cuts, as a function of the total extent of the UV rest-frame emission. {We find that the smallest galaxies show the highest fraction of LAEs}, regardless of the selection. This fraction is $\lesssim10\%$ for galaxies with $r_T^{100}\gtrsim3.5$ kpc and rises to $\gtrsim20\%$ for galaxies with $r_T^{100}\lesssim1.0$ kpc. We compare our results with those reported by \citet{law2012b}, where they find a similar behaviour. We note that they select as LAEs, any galaxy that has a visible Ly$\alpha$ emission, which would translate to an equivalent width cut at 0\AA. Interestingly we do find a {small increase in the fraction of galaxies with \lya\ emission around $r\sim5-6$kpc (with respect to the two adjoining bins)}, similar to what is reported by \citet{law2012}. However, due to the small number of galaxies at these sizes it is likely that this bump is simply a statistical anomaly.

{The results shown in Fig. \ref{fig:lae_fraction_size} are correlated with what we observe in Fig. \ref{fig:class_sizes} concerning the UV sizes for each class. Extended emitters have the smallest UV sizes and are also the ones with the strongest line equivalent widths for Ly$\alpha$}. Our results confirm the trend of smaller UV sizes for larger Ly$\alpha$ equivalent width galaxies reported in the literature \citep[e.g.][]{vanzella2009,law2012b,paulino-afonso2017}. {The prevalence of strong emitters at small UV sizes} can also partially explain the difference in sizes of LAEs and LBGs or regular star-forming galaxies seen by recent studies at $z<4$ \citep[e.g.][]{malhotra2012,wisotzki2016,paulino-afonso2017}  as being due to the nature of LAE selection.

\section{The physical nature of Ly$\alpha$ emitters}\label{sec:discussion}

We have defined a new classification system based on the extent of the Ly$\alpha$ emission as compared to that of the UV continuum. We carried out our visual classification for a sample of $\sim4192$ objects with $2<z\lesssim6$ on the VUDS survey and compiled a sample of 914 galaxies with Ly$\alpha$ in emission and 1919 with no emission line that have been identified in the same class by two different team members. We find that the most common case for line emitters is having coincident line and UV continuum emission ($\sim45\%$) followed by extended line emitters ($\sim24\%$) and then offset line emitters ($\sim11\%$). Other emitters have either a too faint continuum or line to be classified.

We find that extended line emitters are the least massive, least star-forming and have the lowest amount of dust extinction {with respect to the other} main classes. At the same time, it is the class for which the line equivalent width is the largest and the only one with continuum detection for which the stacked spectrum has an equivalent width high enough to be classified as a classical LAE. They also amount to $\sim$43\% of all of the line emitting galaxies with Ly$\alpha$ equivalent width $<-25$\AA\ making them the most likely population (along with single line emitters which amount to $\sim21\%$) to be detected with narrow band surveys. {These extended emitters are also the most compact, show the strongest C{\sc iii}] emission of all of the classes and have the largest \lya\ escape fraction, which are indications of a hard ionizing spectra. This hints that extended emitters are those more likely to be analogues of galaxies in the re-ionization epoch.}

In terms of the physical properties of LAEs, we may expect that part of the {observed} differences is explained with the physical conditions that render their detection possible {in different kinds of surveys}. Selecting stronger equivalent width galaxies (as done in all narrow band surveys) produces a bias towards galaxies with bright extended Ly$\alpha$ emission which we showed are the class of galaxies that have the strongest differences among our objects \citep[similar to those already reported by][]{hathi2016}. {On the other hand,} our sample is primarily comprised of UV-selected objects and thus the observed differences are not as large as pure line surveys which can probe down to lower masses \citep[e.g.][]{gawiser2007,finkelstein2007,guaita2011,vargas2014}. {We can argue that the differences among different studies are most likely due to the difference of continuum selection and line selection in the corresponding surveys, and that if we select galaxies based on optical emission lines (tracing e.g. SFR), both line emitter populations (Ly$\alpha$ and optical) should show similar properties \citep[e.g.][]{hagen2016}.}

For what concerns the derived trend that larger galaxies have a lower fraction of Ly$\alpha$ emitters, there is a possibility of biased measurements due to slit losses. This may happen since  \citet{wisotzki2016} found that the Ly$\alpha$ extension scales with UV continuum size \citep[see also e.g.][]{leclercq2017,yang2017} and therefore, larger galaxies are more likely to have extended Ly$\alpha$ emission beyond the 1\arcsec\ aperture of the slits. In our sample we use an equivalent width cut at $W_\mathrm{Ly\alpha}<-25$\AA\ which, as stated before, includes mostly extended and single line emitters. In turn, these galaxies are the ones with the smallest UV continuum sizes (see Fig. \ref{fig:class_sizes}). {Therefore we are confident that the trend we observe is real and not affected by slit loss problems since most of our strong emitters should have smaller UV sizes and consequently smaller Ly$\alpha$ extension which would reduce the impact of such flux losses. It is possible that a differential surface brightness (SB) dimming effect, which would not affect the UV sizes (measured from deep imaging data) but affect $W_\mathrm{Ly\alpha}$ (as measured from the spectra), which could produce a similar trend. However, we do not expect that to be the case since we expect a core \lya\ emission that has a similar profile shape as the UV continuum, and not a shallower profile which would more easily be affected by SB dimming \citep[][]{leclercq2017}. Also, the lower spatial resolution of VUDS 2D spectra also helps counterbalance any SB dimming effect, since in this case the  SB per pixel is boosted by including photons from a larger physical area.} 

Our findings that galaxies with Ly$\alpha$ emission are smaller than their non-emitting counterparts are consistent with most studies in the literature \cite[e.g][]{malhotra2012,paulino-afonso2017}. However, we are able to show that there is a smooth trend from large to small galaxies where the fraction of strong line emitters increases \citep[][]{law2012b}. And this trend seems to be independent of the  Ly$\alpha$ equivalent width cut we impose, with smaller galaxies being more likely to show Ly$\alpha$ emission.

\subsection{Offset Ly$\alpha$ emission}

There is a possibility that the offset emission (observed in $\sim$10\% of our line emitters' sample) comes from a source in the foreground/background of the UV continuum source. However, there are two strong arguments against {this possibility. One is the fact that the typical offset is $\sim$0.15\arcsec (see  Sect.  \ref{ssection:offsets}) and the probability} of a serendipitous emission occurring at such distances is much smaller than the fraction of offset emitters we find. We expect to find only $\sim0.02(0.05)$ LAEs per arcsec\textsuperscript{2} brighter than $\log_{10}(F_\mathrm{Ly\alpha})=-17.5(-18)$ assuming the detected number counts by \citet{bacon2015}. This would correspond to {the} detection of $\sim$2(5) serendipitous galaxies within a 0.15\arcsec radius for a sample of 4192 galaxies. This is an upper limit on the detection since we can only measure offsets along one spatial direction (East-West of the primary target). By imposing a minimum offset of 0.05\arcsec along that spatial direction the number of expected detections drops to 1(3) Ly$\alpha$. The other is the stacked spectrum for this offset class (see Fig. \ref{fig:class_stacks}). If the detection was not at the same redshift as the continuum source for the majority of the offset sample, we would not detect other lines as the random nature of serendipitous detections would wash out all other lines from the stack.

{
Another possibility is that we are seeing emission from one  Ly$\alpha$-bright clump offset from the galaxy center and produced by violent disk instabilities (VDI). Such instabilities are common in gas-rich high redshift galaxies where high surface densities lead to fragmentation under their own self-gravity to originate massive star-forming clumps \citep[e.g.][]{elmegreen2004,elmegreen2007,genzel2006,genzel2008,dekel2009,dekel2013,bournaud2016}.

Finally, it is also possible that these offsets are indicative of merging systems for which a component is emitting Ly$\alpha$, while the other is not. From the fraction of mergers reported in the literature \citep[e.g.][]{tasca2014,ribeiro2017,ventou2017} we expect that among a sample of 914 galaxies, $\sim20$\% would be involved in major merging. {Nonetheless, we expect major merging to be a sub-dominant process since these offset emitters would need to be a combination of a non-emitting galaxy merging with a single line emitter (and thus a faint, likely low mass object) to match our definition. This is also bolstered by} the distribution of separations between the Ly$\alpha$ emission and the continuum, {where most offsets are relatively small and not compared what one would expect for a major merging event. Thus, to explain the observed fraction of offset emitters as due to merging events, we would assume them to be of minor nature.}

{The combination of multiple processes driving galaxy evolution at high-redshift is consistent with the findings by \citet[][]{ribeiro2017} for the same parent sample. Such a combination of different mechanisms would likely produce two over-imposed distributions of offset distances. However, one would need a higher sample of offset \lya\ emitters to confidently constrain their existence and properties of each individual distribution. Therefore, the large spread in observed offsets in combination with the likely double Gaussian nature of its distribution (see Fig. \ref{fig:offset_dist}) is compatible with the scenario where both disk instabilities and mergers are responsible for the observed offset distances distribution. 
} 


\section{Summary}\label{sec:summary}

The results presented in this paper can be summarized as follows:
\begin{itemize}
\item We have defined four different main classes of star-forming galaxies based on the spatial properties of the Ly$\alpha$ emission as compared to the UV continuum emission. We find a total of 914 galaxies with line emission of which 45\%, 24\% and 11\% have coincident, extended and offset line emission, respectively. We have also a sample of 1918 galaxies with no emission line at all.
\item Extended line emitters show the strongest equivalent widths. In fact, when considering all galaxies with $W_\mathrm{Ly\alpha}<-25$\AA, 41\% of them are extended and $\sim20\%$ are single line emitters. These are, by definition, the most common populations likely to be selected as LAEs in narrow-band imaging surveys.
\item In terms of physical properties, extended line emitters distinguish themselves from the other classes by being the least star-forming,  with lower dust content and smaller UV continuum sizes. {The extended emitters also show the strongest C{\sc iii}] emission and have likely the higher \lya\ escape fraction of all classes.}
\item {We find that the strength of the low ionization ISM lines is correlated with the value of $W_\mathrm{Ly\alpha}$ of each class, with stronger absorptions for galaxies with no \lya\ emission.}
\item We find that the UV size of galaxies correlates with the presence of Ly$\alpha$ emission, with larger galaxies having a lower fraction of LAEs.
\item {From the three main classes of LAGs, offset emitters show the largest velocity differences between the \lya\ emission and the low ionization ISM lines, driven mostly by a larger velocity offset of \lya\ with respect to the systemic velocity. }
\item We find $\sim$11\% of galaxies with offset Ly$\alpha$ emission, {with a median separation between the emission and the UV continum of $1.1^{+1.3}_{-0.8}$ kpc, and a likely double log-normal nature in the distribution of offset distances}. This can be interpreted as a combination of on-going minor merger events or Ly$\alpha$-bright clumps within large galaxies.
\end{itemize}

We conclude that the extent of the line emission is linked to the physical properties of galaxies and that the largest Ly$\alpha$ extent is likely coming from smaller, low-mass galaxies with little dust. Further investigation into this problem is required to confirm the extended nature of Ly$\alpha$ emission in a larger sample by probing down to fainter surface brightness limits. It is likely that future and ongoing observations from large and deep spectroscopic surveys like VANDELS \citep[][]{mclure2018,pentericci2018} and Integral Field Unit observations from MUSE/KMOS/JWST {will help us understand more about the nature of \lya\ emission. Such surveys will place definitive constraints on the observational biases associated with the different classes presented here and will allow for the understanding of the underlying physical motivation for the differences that we report.}



\begin{acknowledgements}
This paper is dedicated to the life of Olivier Le F\`evre: a mentor both in science and in life. The first author will forever be grateful for the opportunity to pursue his scientific curiosity within such great collaboration led by him.
This work is supported by funding from the European Research Council Advanced Grant ERC--2010--AdG--268107--EARLY and by INAF Grants PRIN 2010, PRIN 2012 and PICS 2013.
This work is based on data products made available at the CESAM data center, Laboratoire d'Astrophysique de Marseille. 
This work partly uses observations obtained with MegaPrime/MegaCam, a joint project of CFHT and CEA/DAPNIA, at the Canada-France-Hawaii Telescope (CFHT) which is operated by the National Research Council (NRC) of Canada, the Institut National des Sciences de l'Univers of the Centre National de la Recherche Scientifique (CNRS) of France, and the University of Hawaii. 
This work is based in part on data products produced at TERAPIX and the Canadian Astronomy Data Centre as part of the Canada--France--Hawaii Telescope Legacy Survey, a collaborative project of NRC and CNRS. 
APA, PhD::SPACE fellow, acknowledges support from Fundação para a Ciência e a Tecnologia through the fellowship PD/BD/52706/2014 and the research grant UID/FIS/04434/2013. PC acknowledges support from the BIRD 2018 research grant from the Universit\`a degli Studi di Padova.
This research made use of Astropy, a community-developed core Python package for Astronomy \citep{astropy2013}; Numpy \& Scipy \citep{numpy_scipy}; an Matplotlib \citep{Hunter:2007}.
\end{acknowledgements}


\bibliographystyle{aa}
\bibliography{refs_lya}


\begin{appendix}

\section{Composite spectra for all classes}\label{app:spectra}

This appendix summarizes the information on the composite spectra obtained for each class and the results from line fitting each spectra  (see tables \ref{tab:eqwidth} and \ref{tab:voffset}).

\begin{table*}
\centering
\begin{tabular}{ccccccc}
Line & None & Faint & Coincident & Offset & Extended\\
\hline
Ly$\alpha$ &  $11.40\pm1.22$ &  $-3.57\pm0.26$ &  $-9.75\pm0.36$ &  $-8.12\pm0.34$ &  $-22.41\pm0.57$\\
\hline
SiII1260 &  $1.22\pm0.14$ &  $1.16\pm0.15$ &  $0.91\pm0.13$ &  $0.56\pm0.10$ &  $0.58\pm0.10$\\
OI1302+SII1304 &  $1.79\pm0.18$ &  $1.35\pm0.16$ &  $1.10\pm0.14$ &  $1.83\pm0.17$ &  $1.00\pm0.14$\\
CII1334 &  $1.19\pm0.14$ &  $0.82\pm0.12$ &  $1.07\pm0.13$ &  $0.72\pm0.10$ &  $0.21\pm0.06$\\
SiII1526 &  $0.99\pm0.13$ &  $0.66\pm0.11$ &  $0.56\pm0.10$ &  $0.99\pm0.14$ &  $0.47\pm0.09$\\
\hline
SiIV1397 &  $1.20\pm0.14$ &  $0.53\pm0.10$ &  $1.18\pm0.14$ &  $2.27\pm0.20$ &  $1.27\pm0.15$\\
CIV1549 &  $1.68\pm0.17$ &  $0.81\pm0.12$ &  $0.85\pm0.13$ &  $0.78\pm0.14$ &  $0.82\pm0.11$\\
AlIII1856 &  $0.57\pm0.10$ &  N/A &  $0.85\pm0.12$ &  N/A &  $1.35\pm0.15$\\
\hline
HeII1640 &  $-0.20\pm0.06$ &  N/A &  $-0.32\pm0.08$ &  $-0.94\pm0.12$ &  $-0.72\pm0.11$\\
CIII1909 &  $-0.62\pm0.10$ &  $-1.41\pm0.40$ &  $-1.15\pm0.13$ &  $-0.88\pm0.13$ &  $-2.08\pm0.19$\\
\hline
\end{tabular}
\caption{Equivalent width, in \AA, for features found in the composite spectra. For each line we show the values obtained for the six classes in each of the six columns.}
\label{tab:eqwidth}
\end{table*}

\begin{table*}
\centering
\begin{tabular}{ccccc}
Line & None & Coincident & Offset & Extended\\
\hline
Ly$\alpha$ &  $-632\pm640$ &  $119\pm95$ &  $342\pm123$ &  $-96\pm69$\\
\hline
SiII1260 &  $-144\pm137$ &  $-555\pm161$ &  $-354\pm138$ &  $-400\pm92$\\
OI1302+SII1304 &  $52\pm126$ &  $-339\pm152$ &  $-459\pm136$ &  $-638\pm152$\\
CII1334 &  $-72\pm140$ &  $-248\pm120$ &  $-246\pm118$ &  $-491\pm161$\\
SiII1526 &  $32\pm126$ &  $-175\pm117$ &  $-297\pm151$ &  $-331\pm92$\\
\hline
Average ISM &  $-33\pm132$ &  $-329\pm137$ &  $-339\pm136$ &  $-465\pm124$\\
\end{tabular}
\caption{Velocity offsets, in $\mathrm{km\ s^{-1}}$, for features found in the composite spectra. For each line we show the values obtained for the four main classes in each of the four columns.}
\label{tab:voffset}
\end{table*}

\end{appendix}

\end{document}